\documentclass{cernrepNum}              % CERN Yellow Report: 1up
\usepackage{graphicx,color}       % graphics and colour
\usepackage{epsfig,rotating}      % EPS graphics and rotation
\newcommand{\degc}{$^\circ$C}
\newcommand{\pcc}{cm$^{-3}$}
\newcommand{\pccps}{cm$^{-3}$s$^{-1}$}

\newcommand{\wpm}{Wm$^{-2}$}
\newcommand{\kwpm}{K/Wm$^{-2}$}

\newcommand{\htwoo}{H$_2$O}
\newcommand{\cotwo}{CO$_2$}
\newcommand{\sotwo}{SO$_2$}

\newcommand{\htwosofour}{H$_2$SO$_4$}
\newcommand{\hnothree}{HNO$_3$}

\newcommand{\cft}{$^{14}$C}
\newcommand{\beten}{$^{10}$Be}
\newcommand{\dcf}{$\Delta\,\!^{14}$C}
\newcommand{\doe}{$\delta\,\!^{18}$O}

\newcommand{\etal}{\textit{et al.}}

\newcommand{\asr}{\textit{Adv.\,Sp.\,Res.}}
\newcommand{\atr}{\textit{Atm.\,Res.}}
\newcommand{\bams}{\textit{Bull.\,Amer.\,Meteor.\,Soc.}}
\newcommand{\epsl}{\textit{Earth and Planet.\,Sci.\,Lett.}}
\newcommand{\grl}{\textit{Geophys.\,Res.\,Lett.}}
\newcommand{\jas}{\textit{J.\,Atm.\,Sci.}}
\newcommand{\jastp}{\textit{J.\,Atm.\,Sol.\,Terr.\,Phys.}}
\newcommand{\jgr}{\textit{J.\,Geophys.\,Res.}}
\newcommand{\joc}{\textit{J.\,Clim.}}
\newcommand{\nim}{\textit{Nuc.\,Inst.\,Meth.\,Phys.\,Res.}}
\newcommand{\ptrsa}{\textit{Philos.\,Trans.\,R.\,Soc.\,A}}
\newcommand{\prl}{\textit{Phys.\,Rev.\,Lett.}}
\newcommand{\pnas}{\textit{Proc.\,Nat.\,Acad.\,Sc.\,USA}}
\newcommand{\prsa}{\textit{Proc.\,Roy.\,Soc.\,A}}
\newcommand{\qsr}{\textit{Quat.\,Sci.\,Rev.}}
\newcommand{\ssr}{\textit{Space Sci.\,Rev.}}

\begin{document}

\begin{center}
	{\large EUROPEAN ORGANIZATION FOR NUCLEAR RESEARCH} 
\end{center}
\begin{flushright}
{\large CERN-PH-EP/2008-005} \\ 
{\large 26 March 2008}
\end{flushright}

\begin{center}
\textbf{\large COSMIC RAYS AND CLIMATE}

\vspace{4mm}         % Lower the following text

\textit{\large Jasper Kirkby} \\
{\normalsize CERN, Geneva, Switzerland}

\vspace{12mm}         % Lower the following text
\textbf{\large Abstract}
\end{center}
\begin{center}
\begin{minipage}{140mm} 
Among the most puzzling questions in climate change is that of solar-climate variability, which has attracted the attention of scientists for more than two centuries. Until recently, even the existence of solar-climate variability has been controversial---perhaps because the observations had largely involved correlations between climate and the sunspot cycle that had persisted for only a few decades. Over the last few years, however, diverse reconstructions of past climate change have revealed clear associations with cosmic ray variations recorded in cosmogenic isotope archives, providing persuasive evidence for solar or cosmic ray forcing of the climate. 
However, despite the increasing evidence of its importance, solar-climate variability is likely to remain controversial until a physical mechanism is established.
Although this remains a mystery, observations suggest that cloud cover may be influenced by cosmic rays, which are modulated by the solar wind and, on longer time scales, by the geomagnetic field and by the galactic environment of Earth. 
Two different classes of microphysical mechanisms have been proposed to connect cosmic rays with clouds: firstly, an influence of cosmic rays on the production of cloud condensation nuclei and, secondly, an influence of cosmic rays on the global electrical circuit in the atmosphere and, in turn, on ice nucleation and other cloud microphysical processes. Considerable progress on understanding ion-aerosol-cloud processes has been made in recent years, and the results are suggestive of a physically-plausible link between cosmic rays, clouds and climate. However, a concerted effort is now required to carry out definitive laboratory measurements of the fundamental physical and chemical processes involved, and to evaluate their climatic significance with dedicated field observations and modelling studies. \\[2ex]

\end{minipage}
\end{center}
                                           
\vfill       % Lower the following text

\begin{center}
Published in \\ \textit{Surveys in Geophysics} \textbf{28}, 333--375, doi: 10.1007/s10712-008-9030-6 (2007). \\
The original publication is available at www.springerlink.com
\end{center}

\thispagestyle{empty} % This removes page numbers for current page

% End of cloned EP Preprint cover page

%%%%%%%%%%%%%%%%%%%%%% Adjust page numbering - start %%%%

%\pagestyle{empty}     % Turn off page numbers
\pagestyle{plain}     % Turn on page numbers

\pagenumbering{roman}  % Roman page numbers

\thispagestyle{empty}  % Remove page number for current page

%\newpage \mbox{} \newpage % Add blank page (needs invisible text)
\newpage \tableofcontents \newpage % Produce a table of contents 
 
\pagenumbering{arabic}  % Arabic page numbers 
\setcounter{page}{1}  % Reset the page counter to page 1

%%%%%%%%%%%%%%%%%%%%%% Adjust page numbering - end %%%%

\section{INTRODUCTION}
\label{sec_introduction}

The climate reflects a complex and dynamical interaction between the prevailing states of the atmosphere, oceans, land masses, ice sheets and biosphere, in response to solar insolation. A climate transition involves coupled changes of these systems following an initial perturbation, or ``forcing''. The feedbacks may be sufficiently strong that the net transition is substantially larger or smaller than from the forcing alone. 
Nevertheless, it is important to identify the primary forcing agents since they provide the fundamental reason \textit{why} the climate changed, whereas the feedbacks determine \textit{by how much}.

\textit{Internal} forcing agents (those arising within Earth's climate system) include volcanoes, anthropogenic greenhouse gases and, on very long time scales (tens of millions of years), plate tectonics. Volcanoes have a cooling effect due to increased absorption and reflection of incoming shortwave radiation by aerosols ejected into the stratosphere. Greenhouse gas concentrations determine the absorption and emission of longwave radiation in the atmosphere with respect to altitude. Plate tectonics influence the climate by changing the size, location and vertical profile of land masses, by modifying the air and ocean circulations, and by exchanging \cotwo\ in the atmosphere with carbonates in the lithosphere. 

At present, and ignoring meteor impact, there are only two established \textit{external} forcing agents: orbitally-modulated solar insolation and variations of solar irradiance. Spectral analysis of the glacial cycles reveals precise frequencies that match Earth's orbital variations caused by the gravitational perturbations of the other planets. In fact, the narrow spectral widths---obtained on untuned data---imply that the glacial cycles are driven by an astronomical forcing agent, regardless of the detailed mechanism; oscillations purely internal to Earth's climate system could not maintain such precise phase coherency over millions of years \cite{muller00}. A linkage between orbit and climate is provided by the Milankovitch model, which states that retreats of the northern ice sheets are driven by peaks in northern hemisphere summer insolation.  This has become established as the standard model of the ice ages since it naturally includes spectral components at the orbital modulation frequencies. However, high precision palaeoclimatic data have revealed serious discrepancies with the Milankovitch model that either fundamentally challenge its validity or, at the very least, call for a significant extension \cite{muller00}. 

Variations of solar irradiance by about 0.1\%  have been measured over the 11 y solar cycle, and are well understood in terms of sunspot darkening and faculae brightening \cite{froehlich00}. From the knowledge of stellar evolution and development, the slow progression of irradiance on extremely long time scales (of order tens of millions of years) is also well understood. However, no mechanism has been identified so far for secular irradiance variations on time scales between these two limits. On the other hand, substantial \textit{magnetic} variability of the Sun on decadel, centennial and millennial time scales is well established from magnetometer measurements over the last 150 y \cite{lockwood99} and from archives of  \cft\ in tree rings and \beten\  in ice cores \cite{beer00}. These light radioisotope archives record the galactic cosmic ray (GCR) flux in Earth's atmosphere, which is modulated by the interplanetary magnetic field and its inhomogeneities carried by the solar wind. In earlier studies, long-term solar magnetic variability was assumed to be a proxy for irradiance variability \cite{lean95}. However this assumption lacks a physical basis, and more recent estimates suggest that long-term irradiance changes are probably negligible  \cite{lean02,foukal04,foukal06}.

It is well-established that Earth's climate varies substantially on centennial and millennial time scales \cite{ruddiman01}. Some of these variations may be due to ``unforced'' internal oscillations of the climate system, involving components  with suitably long response times, such as the ice sheets. However, it is hard to explain away all these climate variations as internal oscillations---especially in cases involving large climate shifts (e.g.\,persistent suborbital variations of sea-level by 10--20 m have occurred during both glacial and interglacial climates \cite{thompson05}), or where synchronous climate changes are observed in widely-separated geographical locations, without any clear path for their teleconnection. At present there is no established natural forcing agent on these time scales; they are either too short (solar irradiance, volcanoes)\footnote{Greenhouse gases are not included since, prior to the twentieth century, short-term changes of greenhouse gases, such as occurred during glacial-interglacial transitions, are found to be a feedback of the climate system and not a primary forcing agent \cite{mudelsee01}.} or too long (solar insolation,  plate tectonics). 
In recent years, however, numerous studies of centennial and millennial scale climate change have reported association with GCR variations. 
These are frequently considered in the literature as a proxy for changes of solar irradiance (or a spectral component such as solar ultra violet \cite{haigh03}). The ambiguous interpretation as either a solar-climate or a GCR-climate forcing mechanism can in principle be resolved by examining climate change on different time scales since, unlike solar irradiance, the GCR flux is also affected over longer times by geomagnetic and galactic variations, and over shorter times by solar magnetic disturbances.

The complexity of the climate system means that it is not easy to explain correlations of solar variability and climate in mechanistic models. On the other hand, the observations are too numerous and of too high quality to be ignored or dismissed. The key challenge is to establish a physical mechanism that could link solar or cosmic ray variability with the climate. If a detailed physical mechanism of climatic significance were to be established then the whole field of solar variability would rapidly be transformed into a quantitative and respectable branch of climate science. But what could the cosmic ray-climate mechanism be? Since the energy input of GCRs to the atmosphere is negligible---about $10^{-9}$ of the solar irradiance, or roughly the same as starlight---a substantial amplification mechanism would be required. An important clue may be the reported correlation of GCR flux and low cloud amount, measured by satellite \cite{svensmark97,marsh00,marsh03}. Although the observations are both disputed \cite{kernthaler99, jorgensen00, kristjansson00, kristjansson02, sun02, damon04} and supported 
\cite{usoskin04,harrison06,vieira06}, increased GCR flux appears to be associated with increased low cloud cover. Since low clouds are known to exert a strong net radiative cooling effect on Earth \cite{hartmann93}, this would provide the necessary amplification mechanism---and also the
sign of the effect: increased GCRs should be associated with cooler temperatures. 

Studying past climate variability, before any anthropogenic influence, is a good way to understand natural contributions to present and future climate change. 
If there is good evidence for a significant influence of solar/GCR variability on past climate then it is important to understand the physical mechanism. Despite the lack of evidence for long-term variability of the total or spectral (ultra violet) irradiance, it cannot be ruled out and remains an important candidate.
However since there \textit{is} clear evidence for long-term variability of cosmic rays,  the possibility of a direct influence of cosmic rays on the climate should also be seriously considered \cite{carslaw02}. It is the purpose of this paper to review some of the palaeoclimatic evidence for GCR-climate forcing---on progressively longer time scales---and to consider the possible physical mechanisms and climatic signatures. Finally, we will describe how these mechanisms will be studied under controlled laboratory conditions with the CLOUD facility at CERN.

%---- Begin EPS Fig. (includegraphics)  ----
\begin{figure}[htbp]
  \begin{center}
      \makebox{\includegraphics[width=0.90\textwidth]{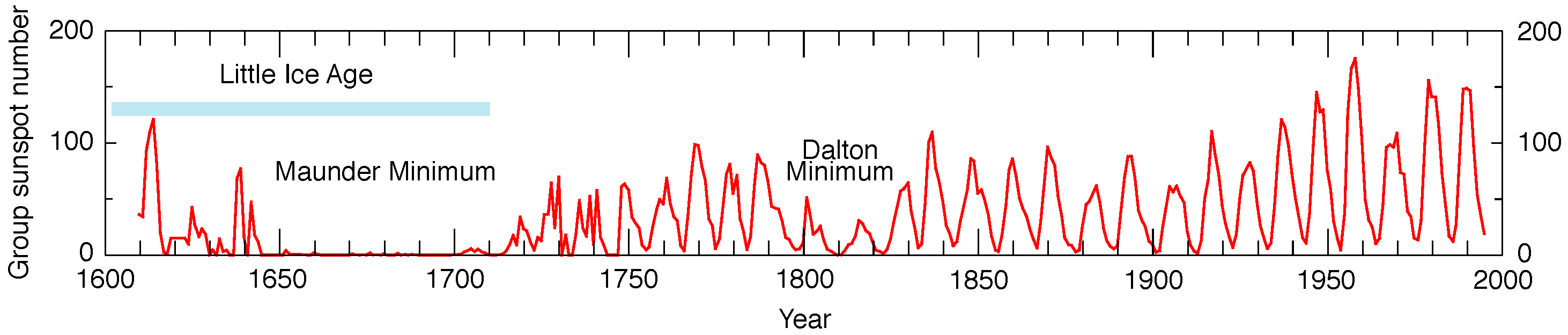}}
%      For a frame: 'makebox' => 'framebox[textwidth]'
  \end{center}
  \vspace{-7mm}
  \caption{Variation of the group sunspot number from 1610 to 1995 \cite{hoyt98}. The record starts 3 years after the invention of the telescope by Lippershey in Holland. The gradual increase of solar magnetic activity since the Maunder Minimum is readily apparent.}
  \label{fig_group_sunspot_record} 
%\end{figure}
%---- End EPS Fig. (includegraphics)  ----
 %---- Begin Fig. (includegraphics)  ----
%\begin{figure}[htbp]
  \begin{center}
\makebox{\includegraphics[height=145mm]{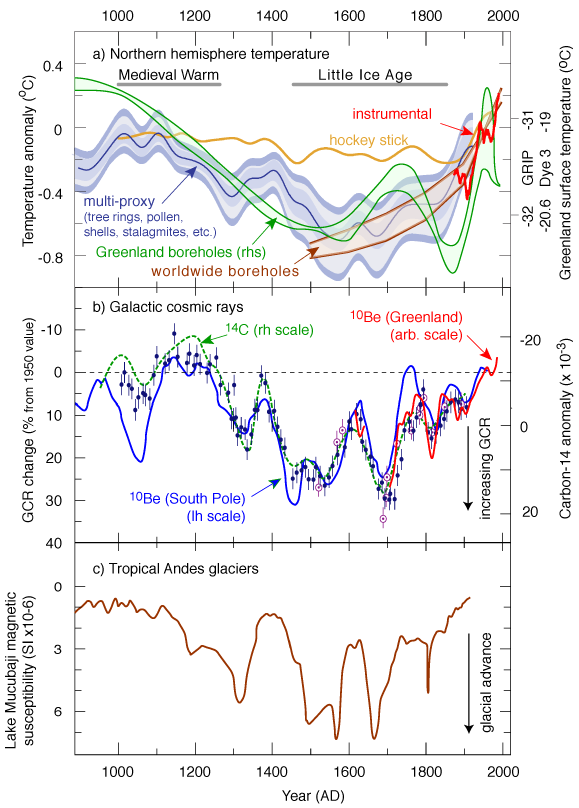}}
%      For a frame: 'makebox' => 'framebox[textwidth]'
%      For a bitmap file: '.png' => '.jpg'
  \end{center}
  \vspace{-7mm}
  \caption{Comparison of variations during the last millennium of a) temperature (with respect to the 1961--1990 average), b) galactic cosmic rays (note the inverted scale; high cosmic ray fluxes are associated with cold temperatures) and c)  glacial advances in the Venezuelan tropical Andes near Lake Mucubaji (8$^\circ$47$^\prime$N, 70$^\circ$50$^\prime$W, 3570~m altitude) \cite{polissar06}. The temperature curves comprise a multi-proxy reconstruction of northern hemisphere temperatures (the band shows 95\% confidence interval) \cite{moberg05}, the so-called hockey-stick curve \cite{mann98,mann99}, borehole temperature measurements worldwide \cite{pollack04} and from Greenland  \cite{dahljensen98}, and smoothed instrumental measurements since 1860. The cosmic ray reconstructions are based on several \cft\ measurements in tree rings (data points and dashed green curve) \cite{stuiver80,klein80}, and \beten\ concentrations in ice cores from the South Pole (solid blue curve) \cite{raisbeck90} and  Greenland (solid red curve) \cite{usoskin02}. The \cft\ anomalies are smaller than those of \beten\ since they are damped by exchanges with the \cotwo\ reservoirs.}  
  \label{fig_gcr_temp_1100y} 
  \end{figure}
%---- End Fig. (includegraphics)  ---- 

\section{SOLAR/COSMIC RAY-CLIMATE VARIABILITY}
\label{sec_solar_gcr_climate_variability}

\subsection{Last millennium}
\label{sec_last_millennium}

\subsubsection{The Little Ice Age and Medieval Warm Period}
\label{sec_lia_mwp}

The most well-known example of solar-climate variability is the
period between 1645 and 1715 known as the Maunder Minimum \cite{eddy76}, during which there was an  almost complete absence of sunspots (Fig.\,\ref{fig_group_sunspot_record}). 
This marked the most pronounced of several prolonged cold spells between about 1450 and 1850 which are collectively known as the Little Ice Age.  During this period the River Thames in London regularly froze across in winter, and fairs complete with swings, sideshows and food stalls were held on the ice. The Little Ice Age was preceded by a mild climate known  as the  Medieval Warm period between about 1000 and 1270. Temperatures during the Medieval Warm period were elevated above normal, allowing the Vikings to colonise Greenland  and wine to be made from grapes in England.

A recent multi-proxy reconstruction of northern hemisphere temperatures \cite{moberg05} estimates that the Little Ice Age was about 0.6\degc\ below the 1961--1990 average, and that climate during the Medieval Warm period was comparable to the 1961--1990 average (Fig.\,\ref{fig_gcr_temp_1100y}a). This contrasts with a widely-quoted earlier reconstruction \cite{mann98,mann99}, known as the hockey stick curve, which indicates a temperature decrease of only 0.2\degc\ between 1000 and 1900, followed by a steep rise in the 20$^{th}$ century (Fig.\,\ref{fig_gcr_temp_1100y}a). However the methodology of the hockey stick analysis has been questioned \cite{mcintyre05}. There have been numerous other temperature reconstructions for the last millennium, which generally lie between these two limits, and their unweighted mean is often used as the combined best estimate \cite{foukal06}. However, this can be misleading since many of the reconstructions share the same temperature proxy datasets, so they are not independent measurements. Perhaps the most accurate measurements of temperatures over the past millennium are provided by geothermal data from boreholes, since these are direct measurements of past temperatures with thermometers---albeit with a modest time resolution that does not preserve rapid changes. The worldwide borehole measurements \cite{pollack04} indicate Little Ice Age temperatures about 0.6\degc\ below those of the mid-20th century, and the Greenland boreholes \cite{dahljensen98} indicate a pronounced Medieval Warm period and Little Ice Age; all favour the multi-proxy reconstruction \cite{moberg05} shown in Fig.\,\ref{fig_gcr_temp_1100y}a.  

\begin{figure}[htbp]
  \begin{center}
      \makebox{\includegraphics[width=120mm]{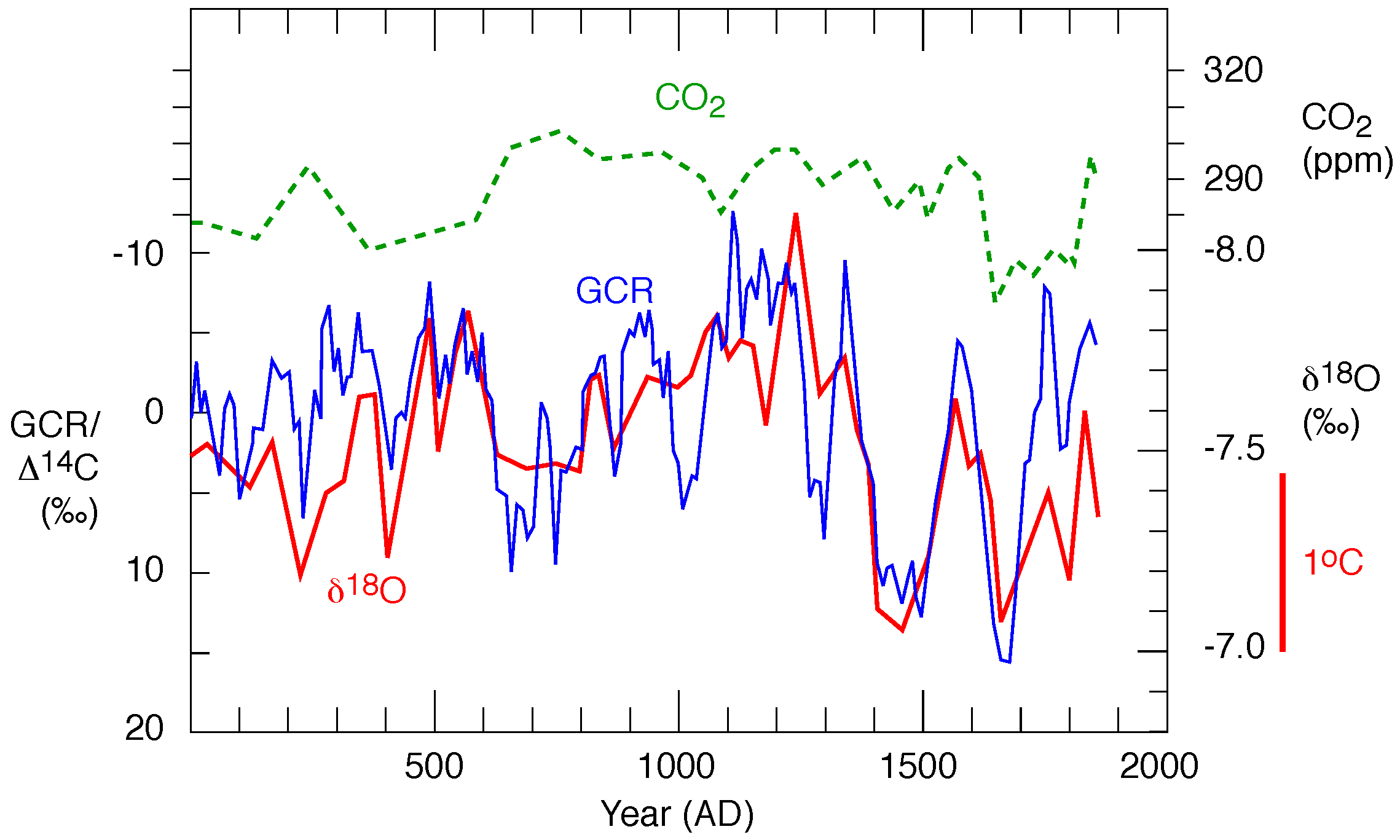}}
%      For a frame: 'makebox' => 'framebox[textwidth]'
%      For a bitmap file: '.png' => '.jpg'
  \end{center}
  \caption{Temperature reconstruction for the Central Alps over the last two millennia, obtained from the \doe\ composition of a speleothem from Spannagel Cave, Austria \cite{mangini05}. The variations of cosmic rays (\dcf) and \cotwo\ over this period are also indicated.}  
  \label{fig_mangini_d18O_2kyr} 
  \end{figure}

For comparison, the variation of GCR intensity over the last millennium is shown in Fig.\,\ref{fig_gcr_temp_1100y}b, as reconstructed from \cft\ in tree rings \cite{klein80}, and \beten\ in ice cores from the South Pole \cite{raisbeck90} and Greenland \cite{usoskin02}. Close similarities are evident between the temperature and GCR records, showing an association of high GCR flux with a cooler climate, and  low GCR flux with a warmer climate. This pattern has been extended over the last two millennia by a reconstruction of Alpine temperatures with a speleothem from Spannagel Cave in Austria (Fig.\,\ref{fig_mangini_d18O_2kyr}) \cite{mangini05}.
Temperature maxima in this region of  central Europe during the Medieval Warm Period were about 1.7\degc\ higher than the minima in the Little Ice Age, and similar to present-day values. The high correlation of the temperature variations to the \dcf\ record (Fig.\,\ref{fig_mangini_d18O_2kyr}) suggests that solar/cosmic ray forcing was a major driver of climate  over this period.

%---- Begin Fig. (includegraphics)  ----
%---- End Fig. (includegraphics)  ----

\subsubsection{Intertropical Convergence Zone}
\label{sec_itcz_influence}

Although the cultural records and temperature reconstructions of the Little Ice Age are predominantly from Europe and the northern hemisphere, numerous palaeoclimatic studies have confirmed the cooling to be a global phenomenon. One example which shows a clear GCR association is a reconstruction of glacial advances in the Venezuelan tropical Andes (Fig.\,\ref{fig_gcr_temp_1100y}c) \cite{polissar06}. The inferred glacial advances  indicate this region experienced a temperature decline of (3.2$\pm$1.4)\degc\ and a 20\% increase of precipitation during the Little Ice Age. 

\begin{figure}[htbp]
  \begin{center}
      \makebox{\includegraphics[width=137mm]{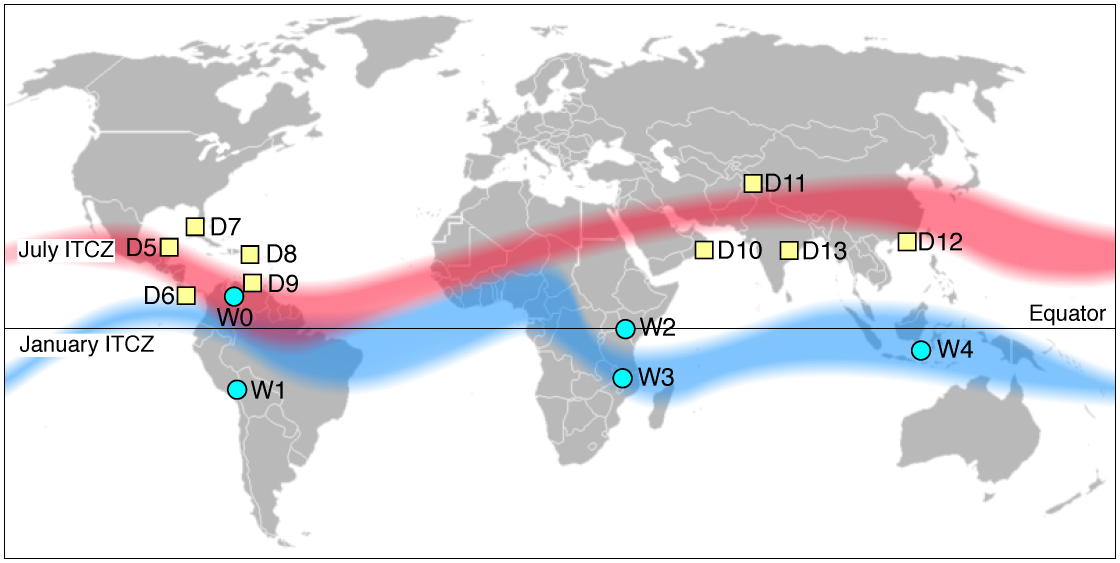}}
%      For a frame: 'makebox' => 'framebox[textwidth]'
%      For a bitmap file: '.png' => '.jpg'
  \end{center}
  \vspace{-5mm}
  \caption{Locations of palaeoclimatic reconstructions of the Little Ice Age indicating wetter (circles labeled W0--W4) or drier (boxes labeled D6--13) conditions than present \cite{newton06}. The approximate most northerly (July) and southerly (January) modern limits of the ITCZ are indicated by the wide transparent bands. The observations indicate a southerly shift of the ITCZ range during the Little Ice Age. The references are W0 \cite{polissar06}, W1 \cite{baker01}, W2 \cite{verschuren00}, W3 \cite{brown05}, W4 \cite{newton06},  D5 \cite{hodell05}, D6 \cite{linsley94}, D7 \cite{lund06}, D8 \cite{wantanabe01}, D9 \cite{haug01}, D10 \cite{anderson02}, D11 \cite{treydte06}, D12 \cite{wang99} and D13 \cite{sinha07}.}  
  \label{fig_itcz_lia} 
  \end{figure}

Another example is a reconstruction of rainfall and drought in equatorial east Africa over the past 1100 y, based on the lake-level and salinity fluctuations of Lake Naivasha, Kenya \cite{verschuren00}. The data show that the climate in this region was significantly drier than today during the Medieval Warm period, and that the Little Ice Age was an extended wet period interrupted by three dry intervals that coincide with the GCR oscillations seen in Fig.\,\ref{fig_gcr_temp_1100y}b.
A similar association of GCRs and tropical rainfall during the Little Ice Age has been observed in the region of the Gulf of Mexico \cite{lund06}. Increased GCR flux seems to have paced increases in the salinity of the Florida Current, leading to a 10\% reduction of the Gulf Stream flow during the Little Ice Age. These examples suggest a possible solar/cosmic ray influence on the Intertropical Convergence Zone (ITCZ). The ITCZ is a region of intense precipitation formed by the convergence of warm, moist air from the north and south tropics by the convective action of the tropical Hadley cells. 
The location of the ITCZ approximately follows the Sun's zenith path, moving north in the northern summer and south in the winter (Fig.\,\ref{fig_itcz_lia}). Numerous other palaeoclimatic reconstructions of the Little Ice Age support the existence of a global influence of GCR flux on tropical rainfall and, moreover, provide the rather clear picture that increased GCR flux is associated with a southerly displacement of the ITCZ (Fig.\,\ref{fig_itcz_lia}) \cite{newton06}. Since the deep convective ITCZ system is the major source of water vapour for the upper troposphere in the tropics and sub-tropics, it controls the supply of the major greenhouse gas (water vapour accounts for about 90\% of Earth's greenhouse effect \cite{freidenreich93}) and the availability of water vapour for cirrus clouds over a large region. The displacement of the ITCZ during the Little Ice Age may therefore imply a substantial global climate forcing.

%---- Begin Fig. (includegraphics)  ----
%---- End Fig. (includegraphics)  ----

\subsubsection{Solar and cosmic ray changes since the Little Ice Age}
\label{sec_solar_cosmic_ray_changes}

\begin{figure}[tbp]
  \begin{center}
      \makebox{\includegraphics[width=85mm]{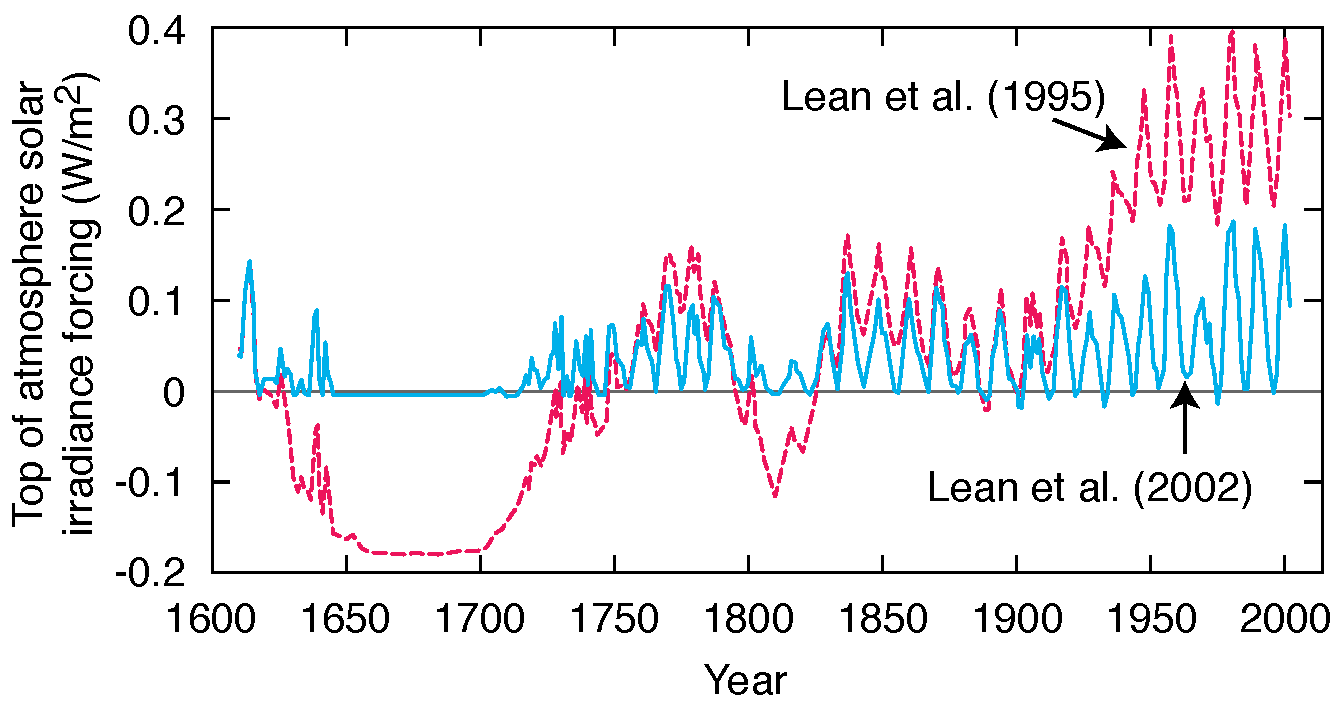}}
%      For a frame: 'makebox' => 'framebox[textwidth]'
%      For a bitmap file: '.png' => '.jpg'
  \end{center}
    \vspace{-5mm}  
  \caption{Top-of-atmosphere solar forcing since 1600 showing previous estimates (dashed curve) \cite{lean95} and present estimates (solid curve) \cite{lean02}.}
  \label{fig_solar_irradiance} 
%  \end{figure}
%---- End Fig. (includegraphics)  ----
%---- Begin Fig. (includegraphics)  ----
%\begin{figure}[tbp]
  \begin{center}
      \makebox{\includegraphics[width=85mm]{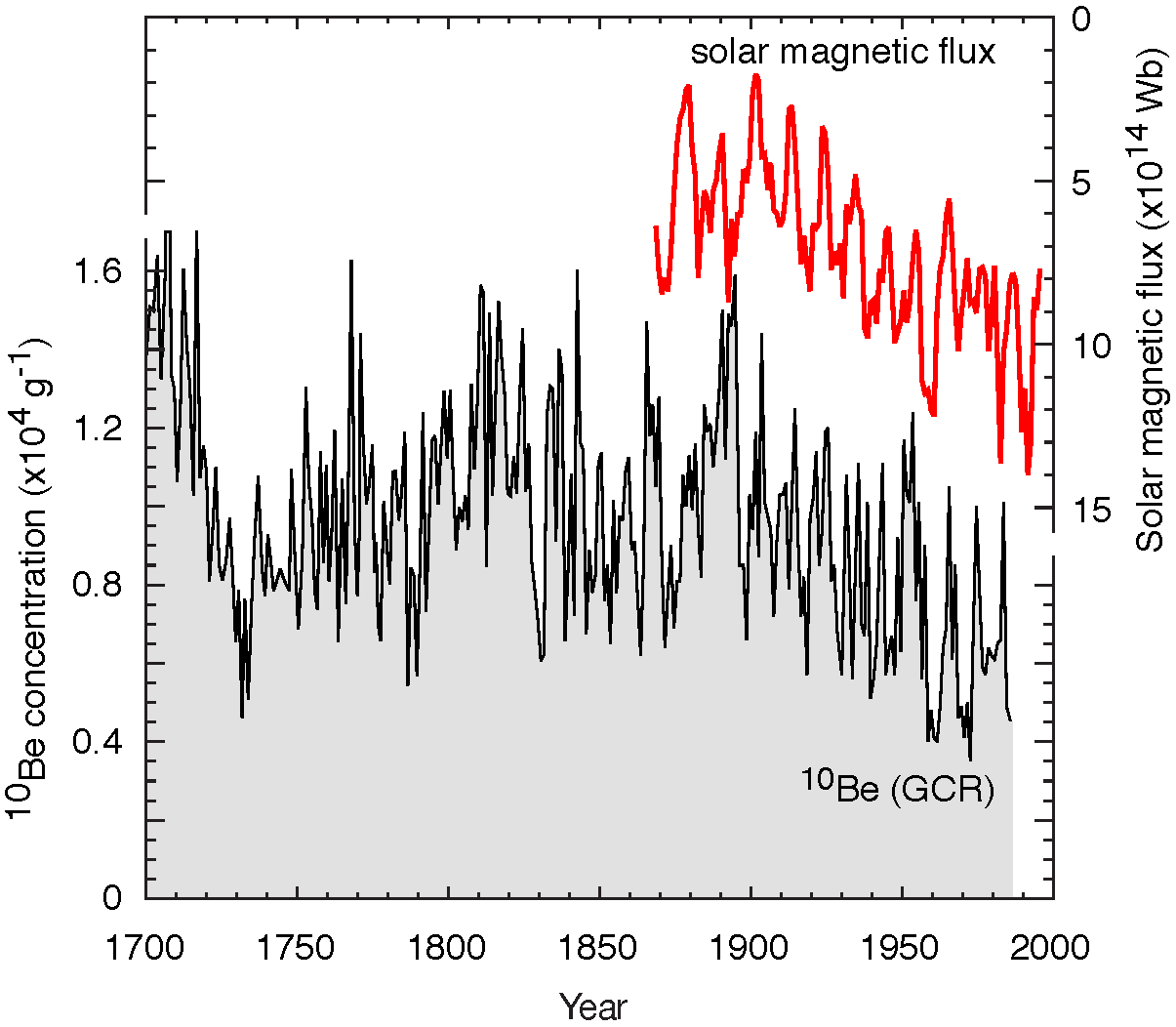}}
%      For a frame: 'makebox' => 'framebox[textwidth]'
%      For a bitmap file: '.png' => '.jpg'
  \end{center}
  \vspace{-5mm}
  \caption{Variations of the \beten\ concentration in Greenland ice cores since 1700 (lower curve) \cite{beer90} and of the solar interplanetary magnetic field since 1870, derived from geomagnetic measurements (upper curve, note inverted scale) \cite{lockwood99}. The \beten\ concentration is directly proportional to the cosmic ray flux integrated over the troposphere and stratosphere. The \beten\ data are unsmoothed, and their short-term fluctuations are dominated by solar cycle modulation.}  
  \label{fig_be10_solarb} 
  \end{figure}

The cold climate of the Little Ice Age appears to have been caused by an extended period of low solar activity. Few sunspots implies low magnetic activity and a corresponding elevated GCR flux. But could the observed warming since the Little Ice Age be explained by changes of solar irradiance rather than introducing the possibility of GCR-climate forcing? Recent results on Sun-like stars, combined with advances in the understanding of solar magnetohydrodynamics \cite{foukal06} have revised earlier estimates of the long-term variation of solar irradiance \cite{lean95} downwards by as much as a factor of five (Fig.\,\ref{fig_solar_irradiance}) \cite{lean02,foukal04}. Apart from the irradiance changes due to sunspot darkening and facula brightening, no mechanism has been identified for solar luminosity variations on centennial or millennial time scales \cite{foukal06}. Current estimates of the secular increase of irradiance since 1700 are therefore based only on the variation in mean sunspot number. The increase in irradiance amounts to less than 0.5~\wpm, which corresponds to about 0.08 \wpm\ at the top of the atmosphere, globally averaged (Fig.\,\ref{fig_solar_irradiance}). Assuming a climate sensitivity of 0.7~\kwpm,  this would contribute less than 0.06\degc\ of the estimated 0.6\degc\ mean global warming between the Maunder Minimum and the middle of last century, before significant anthropogenic contributions could be involved.

%---- Begin Fig. (includegraphics)  ----
%---- End Fig. (includegraphics)  ----

On the other hand, there is clear evidence of a substantial increase in solar magnetic activity since the Little Ice Age (Fig.\,\ref{fig_be10_solarb}) \cite{beer90,lockwood99}. The \beten\ record indicates that the mean global cosmic ray flux has decreased by about 30\% since the the Little Ice Age, with about one half of this decrease occurring in the last century. The good agreement found between the \beten\ and \cft\ records confirms that their variations reflect real changes of the cosmic ray flux and not climatic influences on the transport processes into their respective archives \cite{bard97}.
Around two-thirds of the \beten\ is produced in the stratosphere, and one-third in the troposphere, so the ice cores archives are sensitive to relatively soft primary cosmic rays. 
The higher energy component that penetrates the troposphere has decreased by a smaller amount than suggested by Fig.\,\ref{fig_be10_solarb}. During the twentieth century, the secular reduction of GCR intensity was approximately equivalent to the present-day solar-cycle modulation (Fig.\,\ref{fig_be10_solarb}), corresponding to about 10\% in the troposphere. The decrease is due to an unexplained increase of the solar interplanetary magnetic field by more than a factor of two over this period  (Fig.\,\ref{fig_be10_solarb}) \cite{lockwood99}. In view of the substantial variability of cosmic rays, if they are indeed found to influence cloud cover it could have considerable implications for understanding past climate change and predicting future variability.

%---- Begin Fig. (includegraphics)  ----
\begin{figure}[tbp]
  \begin{center}
      \makebox{\includegraphics[width=100mm]{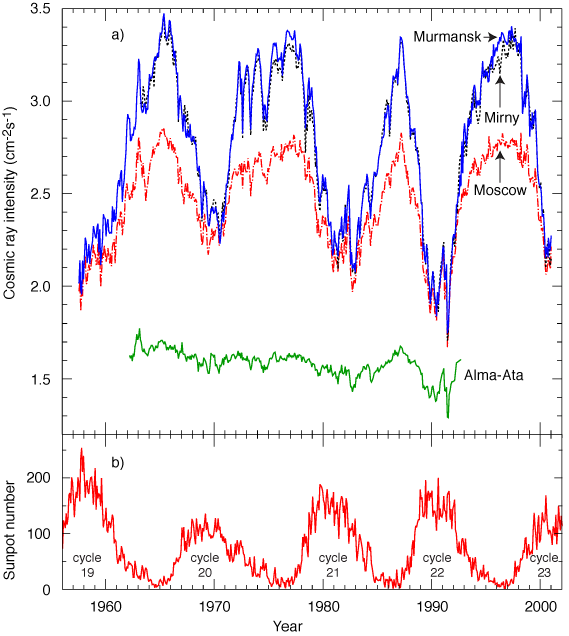}}
%      For a frame: 'makebox' => 'framebox[textwidth]'
%      For a bitmap file: '.png' => '.jpg'
  \end{center}
  \vspace{-5mm}
  \caption{Solar modulation of the galactic cosmic ray intensity for the period 1957--2001: a) balloon measurements of the cosmic ray intensity at shower maximum (15--20~km altitude), measured by the
Lebedev Physical Institute \cite{stozhkov01, babarykin64}, and b) Wolf sunspot number. The curves correspond to four different locations for the balloon flights: Mirny-Antarctica (0.03 GeV/c rigidity cutoff), Murmansk (0.6 GeV/c), Moscow (2.4 GeV/c) and  Alma-Ata (6.7 GeV/c). Due to atmospheric absorption, the data of Murmansk and Mirny practically coincide with each other.}
  \label{fig_gcr_sunspots_lebedev} 
  \end{figure}
%---- End Fig. (includegraphics)  ----

During the second half of the twentieth century, the variation of GCR flux has been directly measured with particle counters. Although these instruments have a low threshold energy (few MeV) for particle detection, the corresponding threshold energy of the primary cosmic rays depends on the type of detected secondary particle and the amount of atmospheric shielding.
In the case of ground-based detectors at sea level, the primary proton threshold energy is about 1.5~GeV for neutron monitors and about 9 GeV for ionization chambers (which detect muons, electrons and protons) \cite{stozhkov01}. The most sensitive measurements are made by balloon-borne ionization chambers, which can reach 30--35~km altitude (4~g/cm$^2$ atmospheric shielding); these have a primary proton threshold of about 0.1~GeV at the highest altitudes. The Lebedev Physical Institute has made continuous measurements of cosmic rays over the last 50 years with balloon-borne detectors at several stations \cite{stozhkov01,babarykin64}. The data for the period 1957--2001 show the solar cycle modulation and the effect of geomagnetic shielding, which leads to reduced fluxes and modulation amplitudes at lower geomagnetic latitudes (Fig.\,\ref{fig_gcr_sunspots_lebedev}). Although the GCR reduction occurred mainly in the first half of the twentieth century (Fig.\,\ref{fig_be10_solarb}), the cosmic ray measurements shown in Fig.\,\ref{fig_gcr_sunspots_lebedev} suggest a continuing decreasing trend in the second half of the century, by a few per cent in the lower stratosphere and upper troposphere.

In summary, the estimated change of solar irradiance between the Little Ice Age and the mid twentieth century is insufficient to explain the observed warming of the climate.  Absence of evidence for a long-term variation of solar irradiance does not, of course, rule it out; precision satellite measurements of solar irradiance have only be available for the last 30 yr. On the other hand, there has been a substantial increase of solar magnetic activity since the Little Ice Age, and a corresponding reduction of the cosmic ray intensity. This suggests that the possibility of an \textit{indirect} solar mechanism due to cosmic-ray forcing of the climate should be seriously considered.

\subsection{Holocene; last 10 ky}
\label{sec_holocene}

\subsubsection{Ice-rafted debris in the North Atlantic Ocean}

Evidence for millennial-scale climate variability during the Holocene has been found in the sediments of ice-rafted debris in the North Atlantic \cite{bond97a,bond97b}.  Deep sea cores reveal layers of foraminifera shells mixed with tiny stones that were frozen into the bases of advancing glaciers and then rafted out to sea by glaciers. These reveal abrupt episodes when cool, ice-bearing waters from the North Atlantic advanced as far south as the latitude of southern Ireland, coincident with changes in the atmospheric circulation recorded in Greenland. These so-called Bond events (named after their discoverer) have occurred with a varying periodicity of 1470$\pm$530~y, during which temperatures dropped and glacial calving increased. The estimated decreases in North Atlantic Ocean surface temperatures are about 2\degc, or 20\% of the full Holocene-to-glacial temperature difference.

\begin{figure}[htbp]
  \begin{center}
      \makebox{\includegraphics[width=110mm]{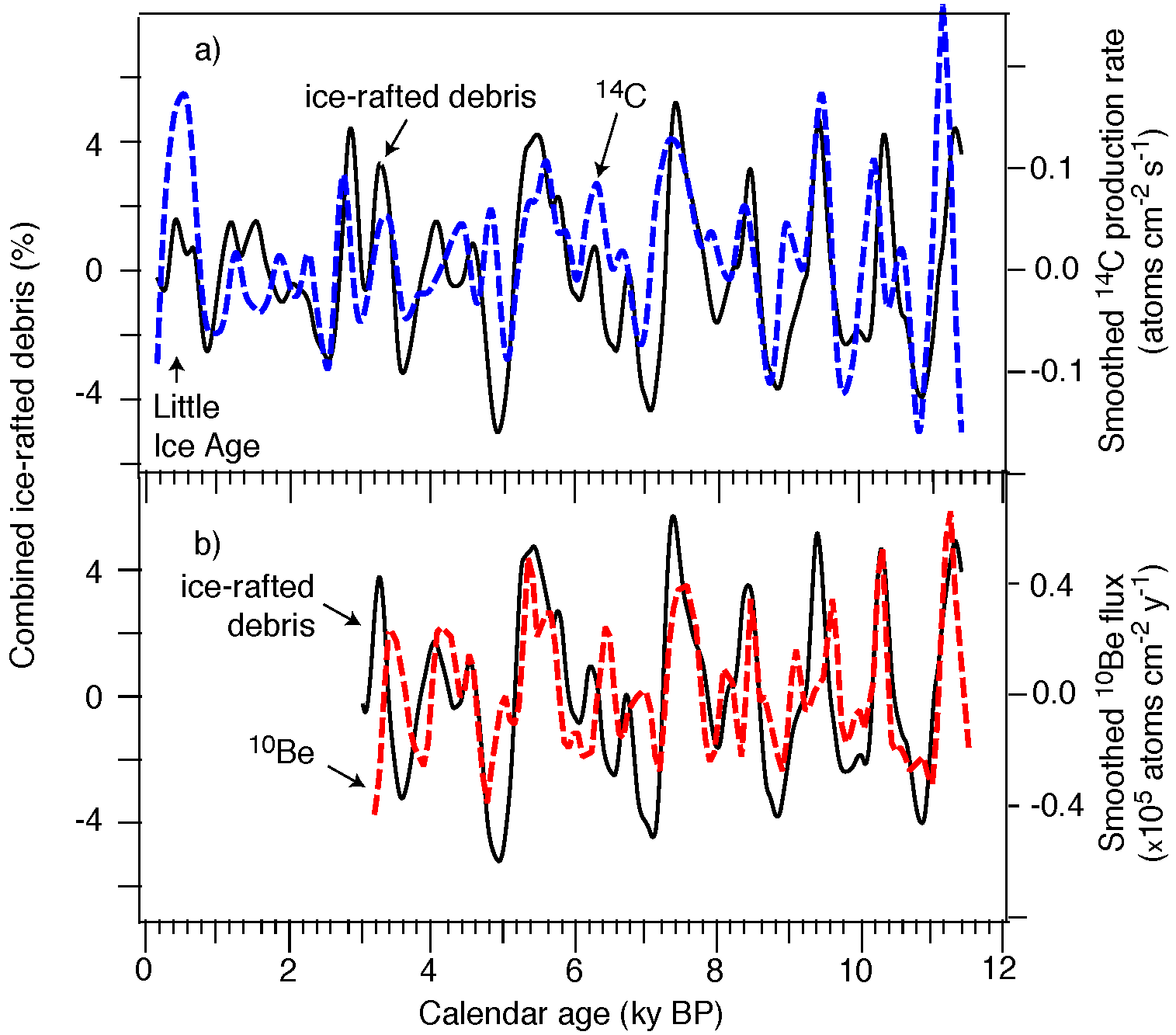}}
%      For a frame: 'makebox' => 'framebox[textwidth]'
%      For a bitmap file: '.png' => '.jpg'
  \end{center}
%  \vspace{-5mm}
  \caption{Correlation of GCR variability with ice-rafted debris
events in the North Atlantic during the Holocene: a) the \cft\ record (correlation coefficient 0.44) and
b) the \beten\ record (0.56), together with the combined
ice-rafted-debris tracers \cite{bond01}.}
  \label{fig_ird_holocene} 
  \end{figure}

What could be the trigger for this millennial-scale climate change?  Orbital variations of insolation are too slow to cause such rapid changes. Ice sheet oscillations are also unlikely to be the forcing agent, for two main reasons. Firstly, the icebergs were launched simultaneously from more than one glacier. Secondly, the events continued with the same quasi-1500 y periodicity for at least the last 30 ky through the Holocene and into the last glacial maximum (but with a larger amount of ice-rafted material)---even though the ice sheet conditions changed dramatically over this interval. Solar/GCR variability appears to be a promising candidate for the forcing agent during the Holocene phase since it is found to be highly correlated with the Bond events (Fig.\,\ref{fig_ird_holocene})  \cite{bond01}. Good agreement is seen between the \cft\ (Fig.\,\ref{fig_ird_holocene}a) and \beten\ (Fig.\,\ref{fig_ird_holocene}b) records, which confirms they are indeed measuring changes of the GCR flux, since their respective transport processes from the atmosphere to archive are completely different. After its formation, \cft\ is rapidly oxidised to $^{14}$CO$_2$ and then enters the carbon cycle and may reach a tree-ring archive. On the other hand, \beten\ attaches to aerosols and eventually settles as rain or snow, where it may become embedded in a stable ice-sheet archive.
The correlation between high GCR flux and cold North Atlantic temperatures embraces the  Little Ice Age, which is seen not as an isolated phenomenon but rather as the most recent of around ten such events during the Holocene. This suggests that the Sun may spend a substantial fraction of time in a magnetically-quiet state.

%---- Begin Fig. (includegraphics)  ----
%---- End Fig. (includegraphics)  ----  

Coincident millennial scale shifts in Holocene climate have now been observed in regions far from the North Atlantic.  In the sub-polar North Pacific, variations of biogenic silica measured in the sediment from Arolik Lake in a tundra region of south-western Alaska reveal cyclic variations in climate and ecosystems over a 10 ky period that are coincident with the Bond events  \cite{hu03}. High GCR flux is associated with low biogenic activity. In equatorial Africa, a 5.4~ky record of rainfall and drought in Lake Edward indicates northward and southward displacement of the ITCZ in phase with the Bond events \cite{russell05}. In southern China, analysis of a stalagmite from Dongge Cave has provided a continuous, high resolution record of the East Asian monsoon over the past 9 ky \cite{wang05}. Although broadly following summer insolation, it is punctuated by 6--8 weak monsoon periods, each lasting 100--500 y, which coincide with the Bond events. A high correlation is found over this entire period between the  Dongge Cave monsoon record (5 y resolution) and the \cft\ record (20 y resolution). 

These results confirm the pattern seen in Fig.\,\ref{fig_itcz_lia} for the Little Ice Age, and extend it throughout the Holocene, namely a high cosmic ray flux is associated with a southerly displacement of the ITCZ. Taken together, the observations suggest that solar/GCR forcing has been responsible for significant centennial and millennial scale climate variability during the entire Holocene, on a global scale.

\subsubsection{Indian Ocean monsoon}

A high resolution record of the Indian Ocean monsoon over the period from 9.6 to 6.2~ky ago has been obtained by analysing the \doe\ composition in the layers of a stalagmite from a cave in Oman \cite{neff01}.  The  \doe\ is measured in calcium carbonate, which was deposited in isotopic equilibrium with the water that flowed at the time of formation of the stalagmite. Oman today has an arid climate and lies beyond the most northerly excursion of the ITCZ, which determines the region of heavy rainfall of the Indian Ocean monsoon system.  Stalagmite growth is evidence that the northern migration of the ITCZ reached higher latitudes at earlier times. In this region, the temperature shifts during the Holocene are estimated to account for only \mbox{0.25 per mil} variation in \doe\ \cite{neff01}, and so the \doe\ variations are mainly due to changes of rainfall.
The data are  shown in Fig.\,\ref{fig_stalagmite_neff}a, together with \dcf\ measured in tree rings such as the California bristlecone pine. The two timescales have been tuned to match bumps within the known experimental errors (smooth shifts have been applied to the U-Th dates up to a maximum of 190 y). During a 430-y period centred around 8.1 ky BP, the stalagmite grew at a rate of 0.55 mm/y---an order of magnitude faster than at other times---which allowed a high resolution \doe\ measurement to be made (Fig.\,\ref{fig_stalagmite_neff}b). 
The striking similarity between the \doe\ and $\Delta\,\!^{14}$C data indicates that that solar/GCR activity tightly controlled the strength of the Indian Ocean monsoon during this 3,000-year period. Increased GCR intensity is associated with a weakening of the monsoon (decreased rainfall) and a southerly shift of the ITCZ, confirming with high time resolution the previous observations .

\subsection{Quaternary; last 3 My}
\label{sec_quaternary}

\subsubsection{Stalagmite growth in Oman and Austria}

There are two primary archives of past GCR changes: the \cft\ record preserved in tree rings, and the \beten\ record in ice cores and ocean sediments. Since \cft\ has a relatively short lifetime ($\tau_{1/2}$ =  5730~y, compared with 1.5~My for \beten), it is only useful for the period up to about about 40~ky ago. In order to determine the atmospheric production rate of these radioisotopes (i.e.\,the GCR flux) from the measured concentrations in a particular climate archive, the transport effects between the atmosphere and the archive must be accounted for. The transport of both \cft\ and \beten\ is strongly influenced by differences between glacial and interglacial climates. Although $^{14}$CO$_2$ is well-mixed in the atmosphere, the concentration is influenced by any changes in ocean ventilation or the extent of sea ice, since the deep ocean is relatively $^{14}$CO$_2$-depleted. In the case of \beten,  changes in atmospheric circulation may sample different latitudinal regions with different \beten\ concentrations. Moreover, the deposition flux depends strongly on the relative rate of wet and dry deposition. For the present climate, the deposition flux of \beten\ in mid-latitude storm tracks is a factor 10 higher than over polar ice sheets, where dry deposition is comparable to (Greenland) or even exceeds (Antarctic) wet deposition \cite{field06}. Extracting the \beten\ atmospheric production rate from the measured flux into the ice sheets therefore has additional uncertainties when large variations occur in the hydrological cycle and in the ice accumulation rates, such as during glacial-interglacial transitions.

%---- Begin Fig. (includegraphics)  ----
\begin{figure}[tbp]
  \begin{center}
      \makebox{\includegraphics[width=100mm]{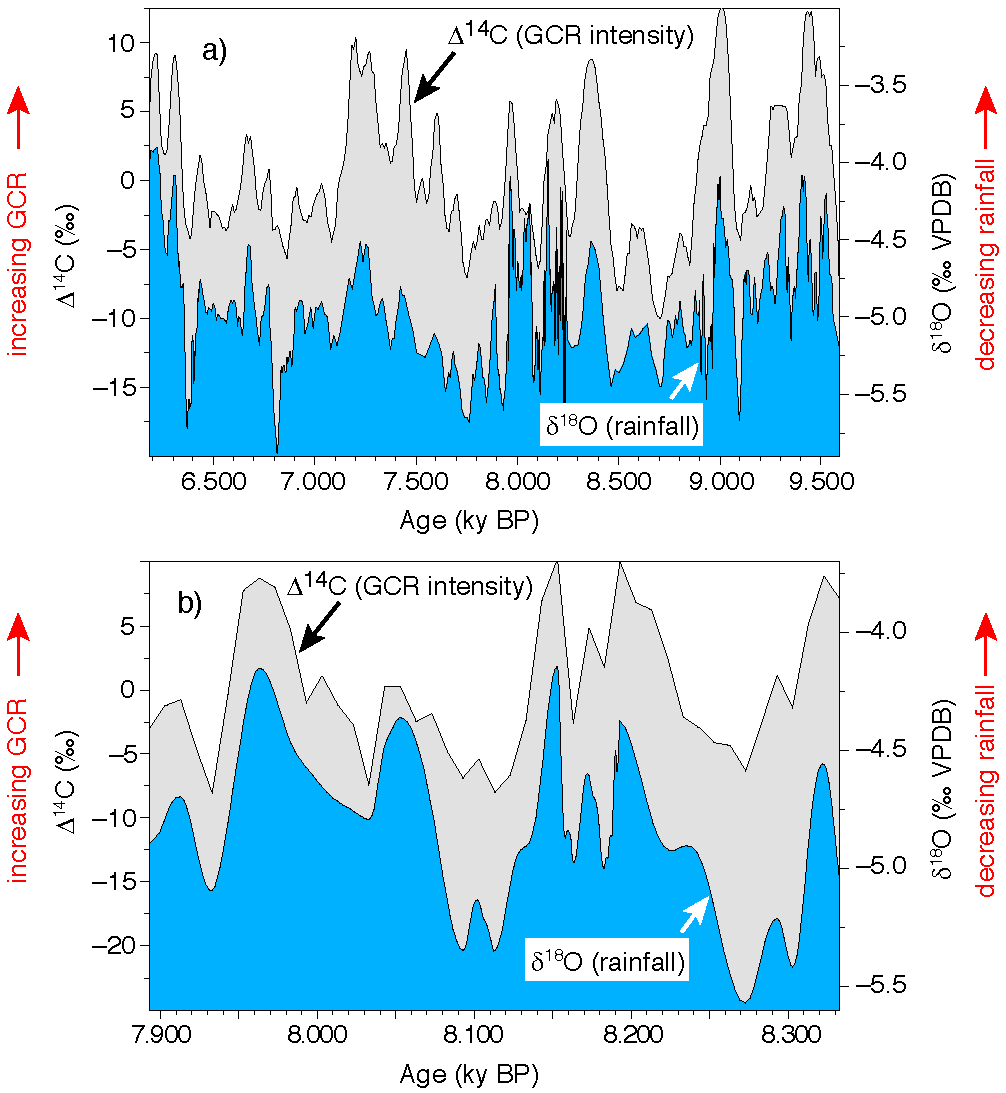}}
%      For a frame: 'makebox' => 'framebox[textwidth]'
%      For a bitmap file: '.png' => '.jpg'
  \end{center}
  \vspace{-5mm}
  \caption{Profiles of \doe\ from a U-Th-dated stalagmite from a cave
in Oman, together with \dcf\ from tree rings in California bristlecone pines and elsewhere, for a) the 3.4 ky period from 9.6 to 6.2~ky BP (before present) and b) the 430 y period from 8.33 to 7.9~ky BP \cite{neff01}. }
  \label{fig_stalagmite_neff} 
  \end{figure}
%---- End Fig. (includegraphics)  ----

\begin{figure}[tbp]
  \begin{center}      
  \makebox{\includegraphics[width=0.85\textwidth]{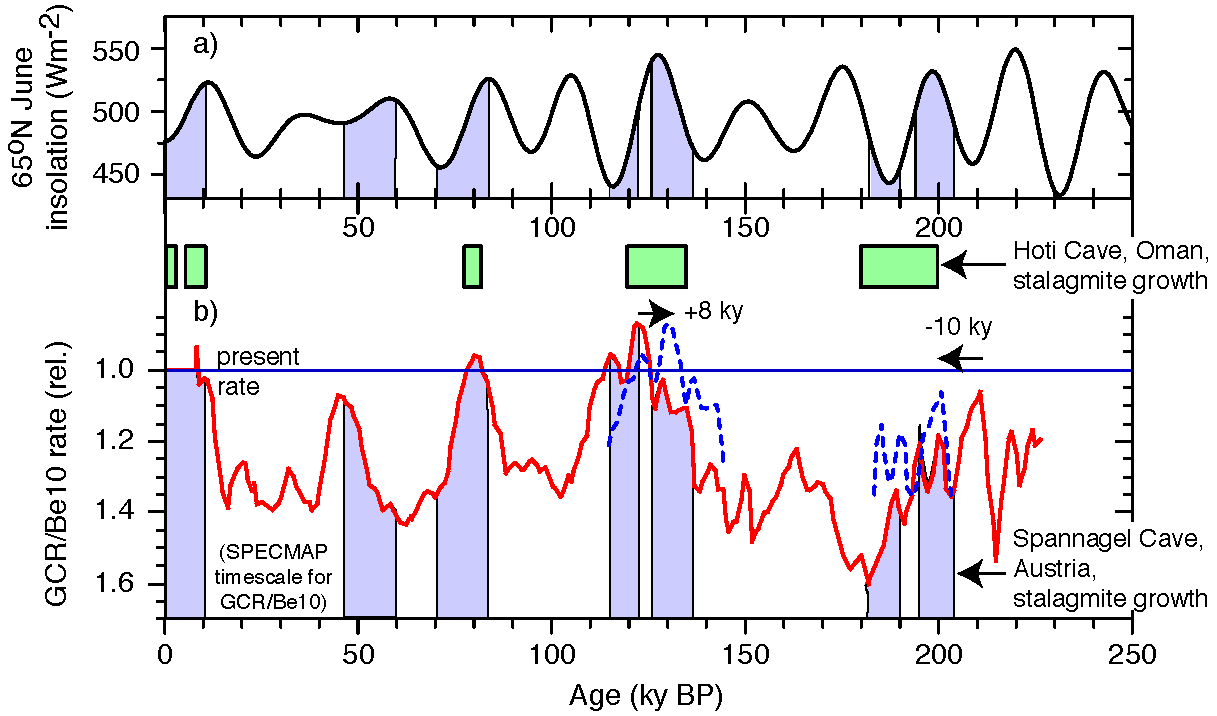}}
%      For a frame: 'makebox' => 'framebox[textwidth]'
%      For a bitmap file: '.png' => '.jpg'
  \end{center}
  \vspace{-5mm}
  \caption{Comparison of the growth periods of stalagmites in Austria and  Oman with a) 65$^\circ$N June insolation and b)~the relative GCR flux (\beten\ ocean sediments; note inverted scale) \cite{christl04}.  The growth periods are indicated by shaded bands (Spannagel cave) or boxes (Hoti cave). Growth periods at Spannagel cave require warm temperatures, close to the present climate; growth periods at Hoti cave require a moist climate. Periods without stalagmite growth are unshaded.  The dashed curves in b) indicate the estimated corrections of systematic errors in the SPECMAP timescale, on which the GCR record is based. The growth periods appear to be associated with intervals of low GCR flux, close to present values.}  
  \label{fig_speleothem_growth} 
  \end{figure}

However, these difficulties are largely avoided in deep ocean sediments, where the \beten\ settling time is about 1 ky, rather than the rapid (1 wk -- 1 y) settling time onto the ice sheets.   Although the long settling time means the ocean archives are insensitive to rapid changes, it has the advantage of a reduced sensitivity to regional variations of production or transport through the atmosphere, ensuring a measurement of global GCR flux. A recent reconstruction of the global \beten\ production rate during the last 220~ky using combined sediment cores from the Pacific, Atlantic and Southern Oceans  shows that the GCR flux was generally around 20--40\% higher than today (Fig.\,\ref{fig_speleothem_growth}b) \cite{christl03,christl04}. This is consistent with palaeointensity reconstructions, both in sign and magnitude, which find an average geomagnetic field strength over this period around  30--40\% weaker than today \cite{guyodo99}. It is also consistent with the observed rise in \cft\ production during the period from 10~ky to 45~ky before present, with roughly equal contributions from the geomagnetic field changes and from changes in the global carbon cycle \cite{beck01}. On the other hand, a recent reconstruction \cite{muscheler04} of the \beten\ flux for the last 50 ky from the Greenland ice core finds no sign of any increase in the \beten\ flux between 10~ky and 30 ky before present, and attributes the discrepancy to the \cft\ data underestimating changes in the carbon cycle---although with large errors that allow for changes of up to 20\% in the production rate (GCR flux).

%---- Begin Fig. (includegraphics)  ----
%---- End Fig. (includegraphics)  ---- 

The ocean core \beten\ data show several 10--15 ky intervals over the last 220~ky when the GCR flux returned to present levels (Fig.\,\ref{fig_speleothem_growth}b). 
These changes are largely due to geomagnetic variability, which is independently measured from palaeointensity data \cite{guyodo99}, although there may also be contributions from solar variability. During the periods of low GCR flux, stalagmite growth was observed in caves located in Oman (Hoti cave) and Austria (Spannagel cave). In contrast, the stalagmite growth periods show no clear pattern of association with 65$^\circ$N June insolation, contrary to the expectations of the standard Milankovitch model (Fig.\,\ref{fig_speleothem_growth}a). The growth periods are precisely dated by U/Th analysis to a $2 \sigma$ precision of 1~ky (Spannagel) or 4~ky (Hoti).  
The growth of stalagmites at each of these locations is especially sensitive to climate variations.  In the case of Oman, water availability requires a warm climate and monsoon rains.  The observed growth periods of the Oman stalagmite extend the Holocene observation shown in Fig.\,\ref{fig_stalagmite_neff} back through the last two glaciations and interglacial---albeit with poorer time resolution---namely,
decreased GCR intensity is associated with a strengthening of the monsoon (increased rainfall) and a northerly shift of the ITCZ in summer. In the case of Austria, Spannagel cave is situated at 2,500 m altitude in the Zillertal Alps, where the present mean air temperature is (1.5$\pm$1)\degc.  Consequently, because of permafrost, stalagmites will not grow at Spannagel if the mean air temperature outside the cave falls by only a small amount below present values.  

Among the serious challenges to the Milankovitch model for the glacial cycles is the timing of Termination II, the penultimate deglaciation. The growth period of stalagmite SPA 52 from Spannagel Cave began at 135$\pm$1.2~ky BP   \cite{spotl02} (Fig.\,\ref{fig_speleothem_growth}b).  So, by that time, temperatures in central Europe were within 1\degc\ or so of the present day.  This corroborates the conclusion, based on sediment cores off the Bahamas, that warming was well underway at 135$\pm$2.5~ky \cite{henderson00}.  Furthermore, dating of a Barbados coral terrace shows that the sea level had risen to within 20\% of its peak value by 135.8$\pm$0.8~ky \cite{gallup02}.  These results confirm the ``early'' timing of Termination II, originally discovered at Devil's Hole Cave, Nevada \cite{winograd92}.  This implies that the warming at the end of the penultimate ice age was underway at the \textit{minimum} of 65$^\circ$N June insolation, and essentially complete about 8~ky prior to the insolation maximum  (Fig.\,\ref{fig_speleothem_growth}a). The Milankovitch model therefore suffers a causality problem: Termination~II precedes its supposed cause. Furthermore, an analysis of deep ocean cores shows that the warming of the tropical Pacific Ocean at Termination II preceded the northern ice sheet melting by 2--3~ky \cite{visser03}. In short, orbital increases of solar insolation on the  northern ice sheets could not have driven the penultimate deglaciation---contrary to the expectations of the Milankovitch  model.    

The \beten\ data, on the other hand, show that the GCR flux began to decrease around 150~ky, and had reached present levels by about 135~ky (Fig.\,\ref{fig_speleothem_growth}b).  This is compatible with Termination II being driven in part by a reduction of GCR flux---although the present \beten\ measurements have large experimental errors. A similar reduction of GCR flux also occurred around 20~ky BP due to a rise in the geomagnetic field strength towards present values.  This coincides with the first signs of warming at the end of the last glaciation, as recorded in the Antarctic ice at about 18 ky BP. Higher-precision \beten\ data from ocean cores, extending to earlier times, are required to investigate this association more closely.

Despite the discrepancies with the Milankovitch model, it remains the only established mechanism that could imprint the orbital frequencies on the glacial cycles. 
At present the GCR measurements from the \beten\ record in ocean sediments have neither sufficient time span nor sufficient precision for spectral analysis and a precise comparison with the orbital frequencies. 
Nevertheless, spectral analysis of the GCR record during the last 220 ky (Fig.\,\ref{fig_speleothem_growth}b) reveals two strong peaks consistent with orbital frequencies, namely 0.01 c/ky (100~ky period) and 0.024 c/ky (41~ky) \cite{kirkby04}. A higher-precision record of the GCR flux over the past few hundred thousand years should in principle be available from the \beten\ record in ice cores, but the challenge here is to disentangle real production rate changes from climate effects on the transport of \beten\ into the ice. There are two possible mechanisms that could modulate GCR ionisation with the orbital frequencies. Firstly, it is speculated that orbital variations could influence the geodynamo, and thereby the GCR flux \cite{malkus68}. Some long-term records of Earth's magnetic field suggest the presence of orbital frequencies, in both strength and magnetic inclination \cite{channell98,yamazaki02}, but others do not \cite{guyodo99}. Secondly, as discussed later in this paper, it is plausible that lightning activity could be modulated by the orbital insolation cycles. If so, this would modulate the ionospheric potential and the atmospheric current density resulting from cosmic-ray ionisation in the atmosphere. 

\subsubsection{Laschamp event}

If cosmic rays are indeed forcing the climate then there should be a climatic response to geomagnetic reversals or excursions (``failed reversal'' events during which the geomagnetic field dips to a low value but returns with the same polarity). A relatively recent excursion was the Laschamp event, when the geomagnetic field fell briefly to around 10\% of its present strength and the global \beten\ production rate approximately doubled \cite{wagner00}. Based on several independent radioisotope measurements, the Laschamp event has been precisely dated at (40.4$\pm$2.0)~ky ago (2$\sigma$ error) \cite{guillou04}.

No evidence of climate change was observed in the GRIP (Greenland) ice core during the Laschamp event \cite{wagner01}. However, this analysis applied a low pass filter to the \beten\ and climate (\doe\ and CH$_4$) measurements, with a cutoff frequency of 1/3000 y$^{-1}$. This is substantially below the estimated 1500~y duration of the Laschamp event \cite{laj00} and has the effect of broadening (diluting) the signal region to 5.5~ky. When the GRIP data are viewed with higher resolution, the Laschamp \beten\ signal coincides with a Dansgaard-Oeschger warming event (both the warming and cooling phases), and so it is difficult to draw a firm conclusion on the absence or presence of an additional GCR-induced climate signal.  

Several climatic effects coincident with the Laschamp event were recorded elsewhere. A pronounced reduction of the East Asia monsoon was registered in Hulu Cave, China \cite{wang01}. At the same time, a brief wet period was recorded in speleothems found in  tropical northeastern Brazil---a region that is presently semi-arid \cite{wang04}. The wet period was precisely U/Th dated to last 700$\pm$400~y from 39.6 to 38.9~ky ago; this was the only recorded period of stalagmite growth in the 30 ky interval from 47 to 16~ky ago. Further evidence has recently been found in deep-sea cores from the South Atlantic, by analysing Nd isotopes as a sensitive proxy of the thermohaline ocean circulation \cite{piotrowski05}. A brief, sharp reduction of North Atlantic deep water (NADW) production was recorded (i.e.\,towards colder conditions), coincident with the Laschamp event. 

This pattern of climate response to the Laschamp excursion of little or no signal at high latitudes (Greenland) but possible signals at low latitudes (China, Brazil, NADW---perhaps reflecting Gulf Stream salinity changes) is consistent with what could be expected from a geomagnetic dip. For a primary cosmic ray to produce ionisation in the lower troposphere, it must create a secondary muon of at least 2~GeV to have sufficient energy to pass through the atmosphere. This implies a primary proton energy of at least about 9 GeV which, for the present geomagnetic field, corresponds to the geomagnetic cutoff at a latitude of about 40$^\circ$. So, for Greenland (which lies above 60$^\circ$N), the cosmic ray flux in the lower troposphere is presently limited by the atmospheric column density and not by the geomagnetic field strength. Even a large reduction of the geomagnetic field---such as during the Laschamp event---would not significantly increase the lower tropospheric ionisation over Greenland (although it would produce a large spike in \beten, as observed \cite{wagner01}, since this reflects the combined tropospheric and global stratospheric ionisation). The regions that would first be affected by a geomagnetic dip are at low latitudes, where the tropospheric ionisation is presently limited by the geomagnetic cutoff; these are indeed the regions that seem to indicate some climate correlations with the Laschamp event. Higher latitudinal regions could be affected later in response to the low-latitude climate changes. However, the short duration of the Laschamp event may not have been sufficient to allow for significant additional cooling to take place at the latitudes of the northern ice sheets, which were already under predominantly-glacial conditions.

 %---- Begin Fig. (includegraphics)  ----
\begin{figure}[htbp]
  \begin{center}
      \makebox{\includegraphics[width=80mm]{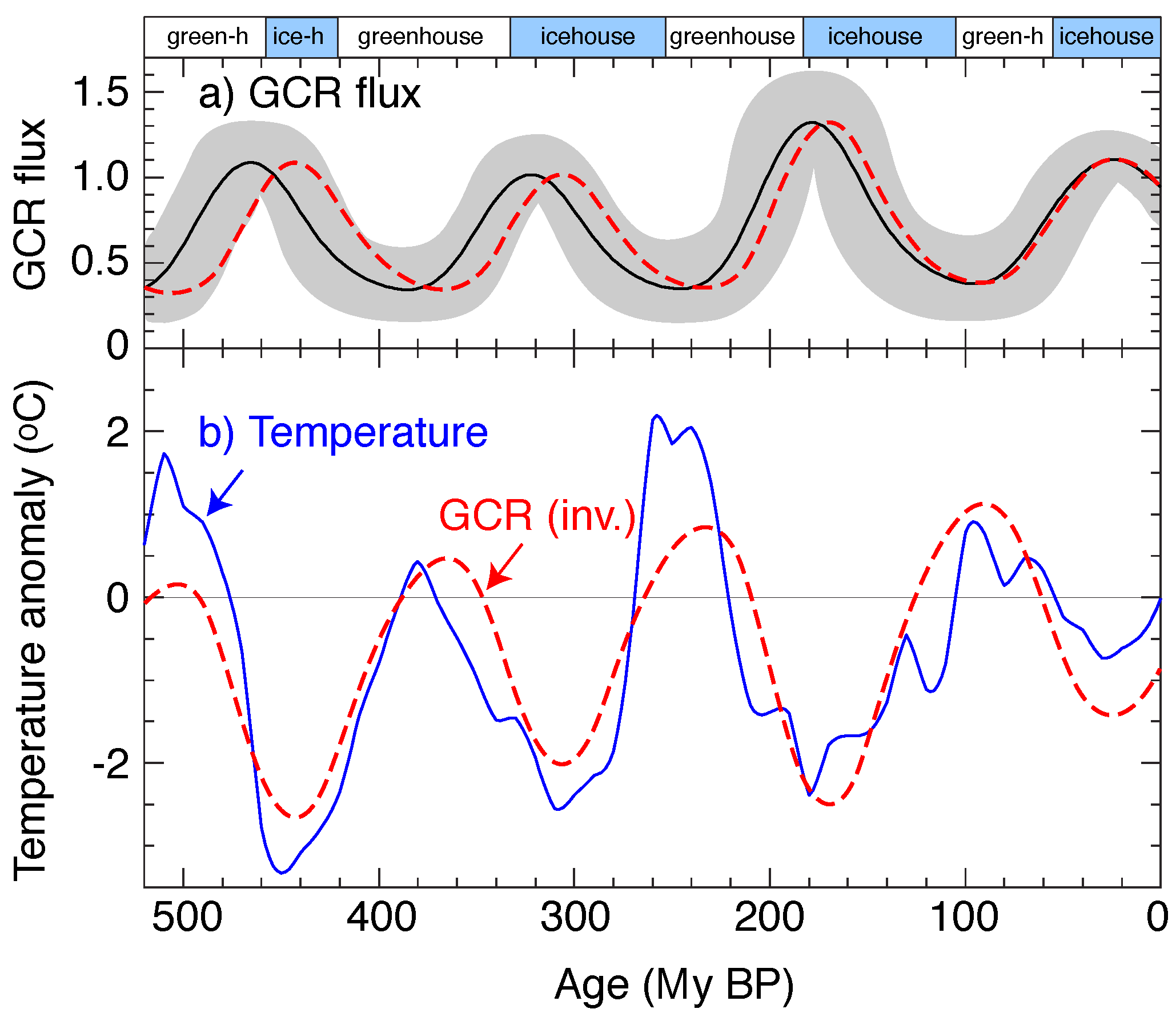}}
%      For a frame: 'makebox' => 'framebox[textwidth]'
%      For a bitmap file: '.png' => '.jpg'
  \end{center}
  \caption{Correlation of cosmic rays and climate over the past 500 My \cite{shaviv03a}: a)~GCR mean flux variations as the solar system passes through the spiral arms of the Milky Way, reconstructed from iron meteorite exposure ages \cite{shaviv02}, and b) ocean temperature anomalies reconstructed from \doe\ in calcite shells found in sediments from the tropical seas \cite{veizer00}. Panel a) shows the nominal reconstructed GCR flux (black curve) and the error range (grey band). The dashed red curves in panels a) and b) shows the best fit of the GCR flux to the temperature data (solid blue curve in panel b), within the allowable error (note inverted GCR scale in panel b). The data are de-trended and smoothed. The dark bars at the top represent cool climate modes for Earth (icehouses) and the light bars are warm modes (greenhouses), as established from sediment analyses elsewhere.}  
  \label{fig_shaviv_veizer} 
  \end{figure}
%---- End Fig. (includegraphics)  ----

\subsection{Phanerozoic; last 550 My}
\label{sec_phanerozoic}

\subsubsection{Celestial cycles}

Tropical sea surface temperatures  throughout the Phanerozoic have been reconstructed from the \doe\ of calcite and aragonite shells found in ocean sediments (30$^\circ$S--30$^\circ$N) \cite{veizer00}. After detrending and smoothing the data, the reconstructed temperatures (Fig.\,\ref{fig_shaviv_veizer}b) show an oscillating behaviour that matches reasonably well with Earth's icehouse and greenhouse climate modes, established from sediment analyses elsewhere (indicated at the top of Fig.\,\ref{fig_shaviv_veizer}).

Independently, a correlation was noted between the occurrence of ice-age epochs on Earth and crossings of the spiral arms of the Milky Way by the solar system, during which elevated GCR fluxes of up to a factor of two are estimated (Fig.\,\ref{fig_shaviv_veizer}a) \cite{shaviv02}. 
The GCR flux is higher within the spiral arms owing to the relative proximity of the supernovae generators and the subsequent diffusive trapping of the energetic charged particles by the interstellar magnetic fields. 
The estimated crossing periodicity of about 140~My is supported by the GCR exposure age recorded in iron meteorites \cite{shaviv02}.  
Both the GCR flux and ocean temperatures appear to follow the same cyclic behaviour and the same phase (Fig.\,\ref{fig_shaviv_veizer}b), whereas estimated \cotwo\ levels do not   \cite{shaviv03a}. 
These observations have been both disputed \cite{rahmstorf04,royer04} and supported \cite{wallmann04,gies05}. However, on quite general grounds, it is hard to conceive of anything other than a galactic mechanism to account for a 140 My periodicity. 

Higher frequency components of the temperature data in Fig.\,\ref{fig_shaviv_veizer}b have also been analysed  \cite{svensmark06a}. 
Significant power is found at 34 My period, and this matches both the estimated phase and  period (34$\pm$6~My) for crossings of the galactic plane by the solar system. This motion corresponds to oscillations above and below the galactic plane with a full cycle of 64--68 My \cite{gies05,svensmark06a}; and it is independent of the motion \textit{in} the galactic plane, as analysed in ref.\,\cite{shaviv02}. 

An alternative mechanism has been proposed to explain these apparently galactic-related periodicities. When passing through the spiral arms of the galaxy, as well as experiencing a higher cosmic ray flux, the solar system encounters giant molecular clouds \cite{frisch00}. These are the remnants of supernovae, and comprise 99\% gas and 1\% dust grains. In extreme cases the clouds may reach densities as high as 2000 H atoms \pcc. This may be compared with about 0.3 H atoms \pcc\ in the present local interstellar cloud which, nevertheless, gives rise to about 40,000 tons of extraterrestrial matter falling on Earth's surface each year. Passage through high density interstellar clouds would cause the heliosphere to collapse below 1 AU, exposing the stratosphere to large quantities of interstellar dust and potentially triggering a ``snowball'' glaciation on Earth by radiative forcing \cite{pavlov05}. Collapse of the heliospheric shield would also lead to a large increase of GCR flux on Earth; the \beten\ rate is estimated to increase up to 400\% of the present value \cite{florinski05}.  Thankfully, present estimates suggest that the frequency of encounter of the solar system with an interstellar cloud of $>$2000 H atoms \pcc\ is rather low (of order one per Gy).

\subsubsection{Biodiversity}

The diversity of life during the course of the Phanerozoic has fluctuated. Analysis of the fluctuations of the number of marine genera has revealed a strong (62$\pm$3)~My cycle and a weaker cycle at 140 My \cite{muller05}. The 140~My cycle may be associated with the GCR-climate cycle, but cold periods in  \doe\ precede the 140~My diversity maxima by 20--25 My. The stronger 62 My cycle does not have any clear origin (a GCR effect due to the vertical oscillation relative to the galactic disk should show up as a 34 My cycle). Using databases covering the Precambrian (4500--543~My ago) and Phanerozoic, an independent measurement of the biodiversity fluctuations has been made from changes of the biological tracer, $\delta^{13}$C  \cite{svensmark06b}. These show a remarkable agreement with the reconstructed GCR flux over a 3.5 By period, based on the expected solar evolution and rate of formation in our galaxy of massive stars that ended life as supernovae. 

An alternative way in which the galactic environment could affect cosmic rays and biodiversity is through isolated events, namely nearby Type IIa supernovae, which could produce a ``cosmic ray winter'' \cite{fields99}. Around one such event is expected per My at a distance of a few $\times$ 10 pc. The first direct evidence for  a close supernova explosion has recently been obtained from an enhancement in the fraction of the $^{60}$Fe radioisotope \mbox{($\tau_{1/2}$ = 1.5~My)} in a deep-sea ferromagnetic crust  \cite{knie04}. The measurements unambiguously show the presence of a nearby supernova that occurred 2.8 My ago; the $^{60}$Fe amounts are compatible with a distance of a few $\times$ 10 pc. This is estimated to have increased the GCR flux by around 15\% for a period of about 100 ky. Under the assumption of GCR-climate forcing, this would have caused a prolonged cold period; coincidentally or not, the onset of northern hemisphere glaciation occurred at the same time.

\section{MECHANISMS}
\label{sec_gcr_climate_mechanisms}

\subsection{GCR-cloud mechanisms}
\label{sec_gcr_cloud}

\subsubsection{GCR characteristics}

Galactic cosmic rays mainly comprise high energy protons which have been accelerated by supernovae and other energetic sources in the Milky Way galaxy. The main flux is in the energy region of a few GeV, although the spectrum extends eleven orders of magnitude higher (albeit with negligible intensity). The primary cosmic rays interact in the atmosphere at around 30 km altitude, producing showers of secondary particles which penetrate the troposphere; below 7 km, the secondaries are mainly muons.  The charged cosmic rays lose energy by ionisation and, away from continental sources of radon,  are responsible for essentially all of the fair-weather ionisation in the troposphere. Ion-pair production rates vary between about \mbox{2~\pccps} at ground level, 10~\pccps\ at 5~km, and 20--50~\pccps\ at 15 km altitude, depending on geomagnetic latitude. Taking into account production rates and loss mechanisms such as recombination and attachment to aerosols, the equilibrium ion pair concentrations vary between about 200--500~\pcc\ at ground level and 1000--3000~\pcc\ in the lower stratosphere \cite{ermakov97}. The GCR flux is modulated (typically by a few tens of per cent, depending on latitude and altitude)  by the solar wind, the geomagnetic field strength  and the galactic environment of the solar system.  Since these vary over periods from tens to billions of years, a putative GCR-climate forcing has the potential to affect climate on all time scales.

Two different classes of mechanisms have been proposed to link the GCR flux with clouds. The first hypothesis is that the ionisation from GCRs influences the production of new aerosol particles in the atmosphere, which then grow and may eventually increase the number of cloud condensation nuclei (CCN), upon which cloud droplets form. The second hypothesis is that GCR ionisation modulates the entire ionosphere-Earth electric current which, in turn, influences cloud properties through charge effects on droplet freezing and other microphysical processes.

\subsubsection{Ion-induced nucleation of new aerosols}
\label{sec_iin}

Aerosols are continuously removed from the atmosphere by wet and dry sedimentation, by self coagulation, and by scavenging in clouds. As a result they have short lifetimes---of order a day---and the rate at which new aerosols are produced can have a large influence on clouds. An important source of new aerosols is thought to be nucleation from trace condensable vapours in atmosphere, but these processes remain poorly understood \cite{kulmala04}. The atmosphere contains very few trace gases that are suitable for nucleation; the most important is thought to be  sulphuric acid, although ammonia and iodine oxides may also be significant. However, the concentration of sulphuric acid in the atmosphere is generally too low to drive the subsequent growth to CCN sizes. Over land masses, organic compounds such as acetone and the oxidation products of terpenes emitted by trees are thought to be important for aerosol growth. In marine regions, isoprene oxidation products and iodine oxides are thought to participate in aerosol growth. 

%---- Begin Fig. (includegraphics)  ----
\begin{figure}[htbp]
  \begin{center}
      \makebox{\includegraphics[width=150mm]{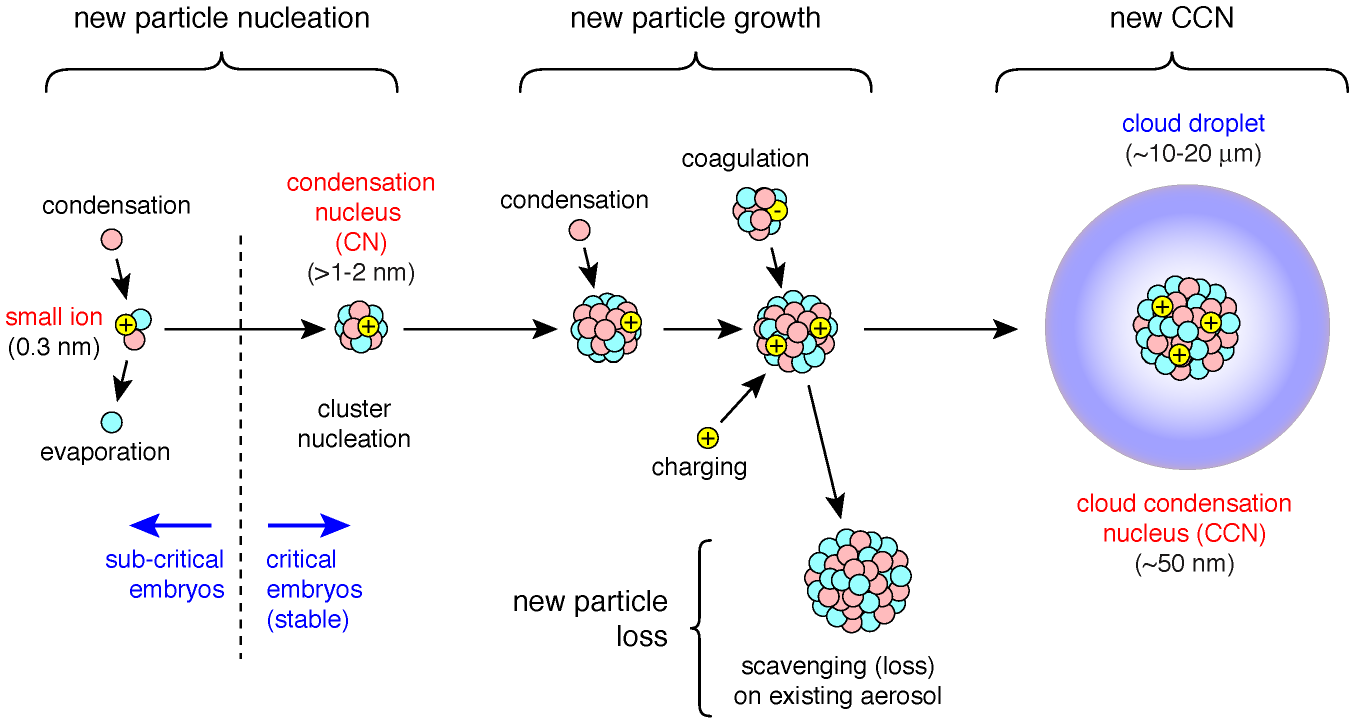}}
%      For a frame: 'makebox' => 'framebox[textwidth]'
%      For a bitmap file: '.png' => '.jpg'
  \end{center}
\vspace{-5mm}
  \caption{Ion-induced nucleation of new particles from trace condensable vapours and water in the atmosphere. }  
  \label{fig_ion_induced_nucleation} 
  \end{figure}
%---- End Fig. (includegraphics)  ----

New aerosols consist of small clusters containing as few as two molecules  (Fig.\,\ref{fig_ion_induced_nucleation}). These may grow by further condensation of molecules of trace vapours and water---or they may evaporate. Above a certain critical size, the cluster is thermodynamically more likely to grow by further condensation; below the critical size, it is more likely to shrink by evaporation. The presence of charge stabilises the embryonic cluster through Coulomb attraction, and reduces the critical size---a process known as ion induced nucleation \cite{yu00,yu01,laakso02}. 

As well as inducing particle formation, charge also accelerates the early growth process, due to an enhanced collision probability, and thereby increases the fraction of particles that survive removal by coagulation before reaching sizes of 5 nm \cite{laakso03}. Beyond this size, electrostatic effects are thought to be negligible, but coagulation rates also fall strongly due to the lower particle mobility. So, with regard to enhancing the survival of new particles to CCN sizes, charge operates precisely in the most critical range, below 5 nm. Nevertheless, only a small fraction of new particles reach the minimum size to be effective CCN (about 50 nm), since most are lost by scavenging on existing aerosols. Geographical regions of very low pre-existing aerosol loading are therefore the most favorable for growth of fresh CCN from new particle production. 

Particle formation in the upper troposphere tends to occur readily since the condensable vapours are often supersaturated (due primarily to the low ambient temperatures and the small pre-existing aerosol surface area). In the boundary layer, however, nucleation tends to occur during ``aerosol bursts'', characterised by long periods without any new production followed by brief periods of rapid production and growth lasting a few hours. These bursts presumably occur when the threshold conditions, such as trace vapour concentrations, are reached so that nucleation and/or growth to detectable sizes \cite{kulmala07} can take place. In ion induced nucleation models, the presence of charge from GCRs lowers this threshold. The observed particle production rates are typically in the range 0.1--10 \pccps, although they may be as high as $10^4$ \pccps\ or more \cite{kulmala04}. In the lower troposphere the ion pair production rate from GCRs is about 5 \pccps\, and the equilibrium small ion concentration is about 500~\pcc, so only some of the nucleation events could be readily explained by ions. A further complication in assessing the contribution of ions to nucleation bursts is that the counting threshold of present instruments is 3~nm aerosol size, whereas the nucleation occurs around 0.5 nm. A large but uncertain fraction of new particles may therefore be lost by coagulation before they are detected, so the true nucleation rate has not yet been measured \cite{dalmaso02}. In addition, a large population of neutral ultrafine aerosol may exist below the 3 nm threshold, whose growth is inhibited due to lack of condensable vapours \cite{kulmala07}. In this case, it is possible that ion induced nucleation could slowly populate the ultrafine aerosols over an extended time, allowing a mechanism for large aerosol bursts that exceed the instantaneous ion-pair production rate from GCRs.  

There is some experimental and observational evidence to support the presence of ion induced nucleation in the atmosphere.
Early studies, beginning in the 1960's, \cite{vohra69, vohra84} demonstrated ultrafine particle production from ions in the laboratory, at ion production rates typically found in the lower atmosphere. This has also been found in a more recent laboratory experiment, under conditions closer to those found in the atmosphere \cite{svensmark06c}. Observations of ion-induced nucleation in the upper troposphere have also been reported \cite{eichkorn02,lee03} and also of aerosol bursts in the lower troposphere \cite{laakso04}, although their rate frequently exceeds what could be caused by instantaneous ion-induced nucleation. Laboratory measurements have shown that ions are indeed capable, under certain conditions, of suppressing or even removing the barrier to nucleation in embryonic molecular clusters of water and sulphuric acid at typical atmospheric concentrations, so that nucleation takes place at a rate simply determined by the collision frequency \cite{froyd03a,froyd03b,lovejoy04}.

\subsubsection{Global electric circuit}
\label{sec_global_electric_circuit}

The second mechanism that may link GCR flux and clouds concerns the effect of GCRs on the global electric current flowing between the ionosphere and Earth's surface \cite{tinsley00a,harrison03}. Except for a contribution from radioactive isotopes near the land surface, GCRs are responsible for generating all the fair-weather atmospheric ionisation between ground level and the mid mesosphere, at about 65 km altitude. As such, GCRs fundamentally underpin the global electrical circuit.

%---- Begin Fig. (includegraphics)  ----
\begin{figure}[htbp]
  \begin{center}
      \makebox{\includegraphics[width=65mm]{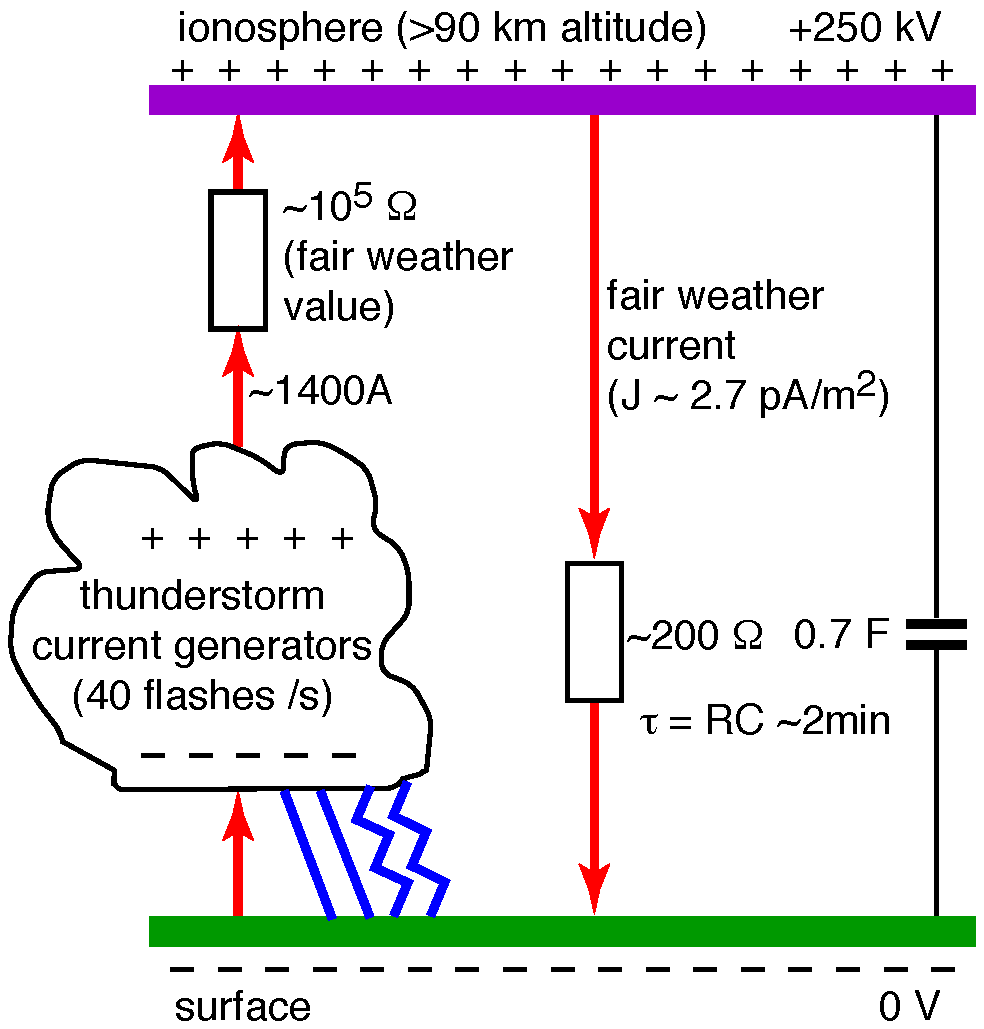}}
%      For a frame: 'makebox' => 'framebox[textwidth]'
%      For a bitmap file: '.png' => '.jpg'
  \end{center}
  \caption{Schematic of the global electrical circuit. The current
generators are thunderstorms, which are located predominantly over tropical land masses. The return path is the global fair-weather current flowing between the ionosphere and ground. It varies between 1 and 4 pA/m$^2$, depending on the local atmospheric columnar resistance, which is mainly determined by the surface altitude and the GCR flux. A continuous current supply is required to maintain the ionospheric potential near +250 kV since the decay time constant of the ionosphere-ground spherical capacitor is only about 2 minutes.}
  \label{fig_global_electric_circuit} 
  \end{figure}
%---- End Fig. (includegraphics)  ----

The atmospheric electric circuit involves a global current of 
about 1400 A (and 400 MW power) which is  sustained by thunderstorms continuously active over tropical land masses (Fig.\,\ref{fig_global_electric_circuit}) \cite{kraakevik61,rycroft00}. The thunderstorms carry negative charge to the ground and an equivalent positive current flows up to the ionosphere. Due to the high currents, large electric fields are generated above thunderstorms and the air can break down electrically to produce optical flashes known as sprites and elves.
This current generator maintains the ionosphere at a relative positive potential of about 250~kV.
The return current between the ionosphere and the Earth's surface flows throughout the atmosphere, in regions of disturbed and undisturbed weather, and is carried by vertical drift of small ions; the average fair-weather vertical current density, $J_z$, is 2.7\,pA\,m$^{-2}$ \cite{kraakevik61,rycroft00}. 
By controlling the lower atmospheric ion pair concentration, GCRs can affect $J_z$, the atmospheric conductivity, the ionospheric potential, and the vertical potential gradient in the atmosphere. In polar regions, these quantities are also influenced by low energy ionising particles from relativistic ($\sim$1 MeV) electron precipitation, which may occur during  crossings of the solar wind magnetic sector boundaries.

Decadel changes in the ionospheric potential and surface potential gradient have in fact been observed and attributed to changes in the GCR flux \cite{markson81, harrison04}. The positive correlation observed between GCR flux and surface potential gradient \cite{harrison04} would imply an influence on the current source in order to sustain the higher net flow of fair-weather charge through the atmosphere, which, at the same time, has a higher conductivity. The simplest explanation is that the GCRs affect the conductivity or electrical breakdown between the top of thunderstorms and the ionosphere; increased GCR flux could then lead to  increased current generation.

In addition to GCR variations, the global electrical circuit is affected by coupling between the magnetosphere and the solar wind. The relative velocity between the magnetosphere and the interplanetary magnetic field (IMF) carried by the solar wind plasma generates additional potential distributions superimposed on the global ionospheric potential. These include a dawn-dusk potential difference of 30--150 kV across each polar cap ionosphere (generated by the north-south component of the solar wind magnetic field, $B_z$), and a potential difference of several tens of kV between the northern polar cap and the southern polar cap ionosphere (generated by the east-west component, $B_y$). Moreover, when $B_z$ points in a southerly direction, the interplanetary magnetic field directly couples with the geomagnetic field and gives rise to intense auroral activity in the polar regions, involving greatly enhanced charge deposition at high altitudes (above 30km, with bremsstrahlung X- and gamma rays constituting the most penetrating radiation).

%---- Begin Fig. (includegraphics)  ----
\begin{figure}[htbp]
  \begin{center}
      \makebox{\includegraphics[width=140mm]{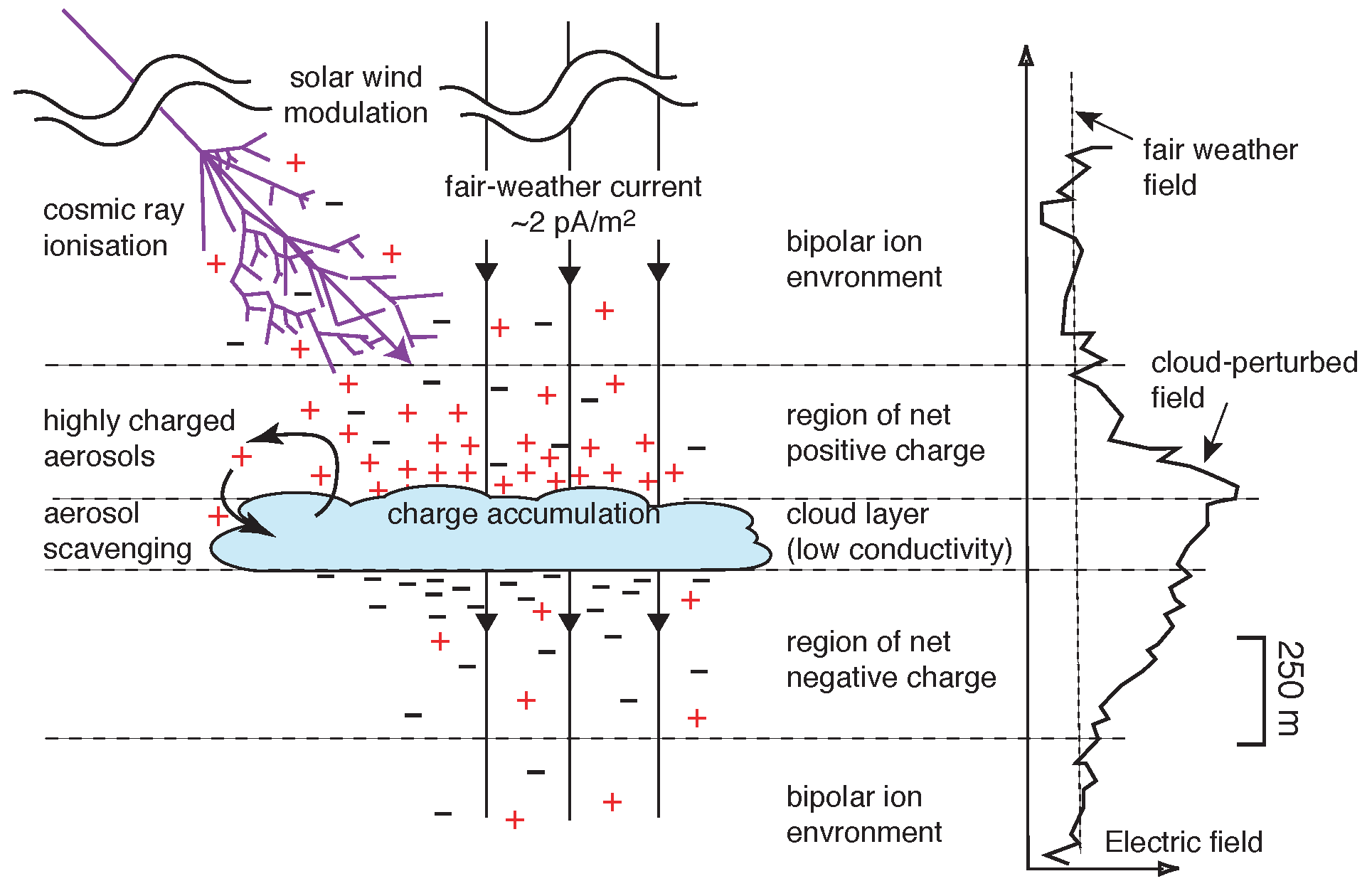}}
%      For a frame: 'makebox' => 'framebox[textwidth]'
%      For a bitmap file: '.png' => '.jpg'
  \end{center}
  \vspace{-5mm}
  \caption{Schematic showing the creation of unipolar space charge at the top and bottom of clouds due to the drift ions from cosmic rays in the fair weather electric field. This space charge can become attached to droplets and aerosol particles, and then entrained within clouds where it may influence microphysical cloud interactions.}  
  \label{fig_charge_scavenging} 
  \end{figure}
%---- End Fig. (includegraphics)  ----

What effect might these changes in the electrical circuit have on the climate? The basic hypothesis is that the vertical conduction current generates highly charged droplets $(\!\gappeq$100 $e$) at the upper and lower boundaries of clouds (Fig.\,\ref{fig_charge_scavenging}) \cite{tinsley00a}. When these droplets evaporate they leave behind highly charged and coated ``evaporation nuclei''. Some of these will be ice nuclei: particles that induce ice formation. Field measurements show that the tops of clouds frequently contain a great deal of supercooled liquid water in the temperature range  -40\degc\ to 0\degc, since ice nuclei are rare in the atmosphere. The highly charged evaporation nuclei may then be entrained inside the clouds, where their charge may enhance collisions with liquid cloud droplets by ``electroscavenging'' \cite{tinsley00b}.  If the charged aerosol is an ice nucleus, this will induce freezing by so-called contact nucleation, which is a particularly efficient freezing mechanism \cite{sastry05}. This would imply that increased GCR intensity leads to increased ice particle formation in clouds. The formation of ice crystals in clouds strongly influences cloud precipitation since ice crystals grow rapidly in the presence of liquid water. Furthermore, the release of latent heat can influence atmospheric dynamics.

In addition to droplet freezing, high electric charges will affect the microphysics of the liquid-phase droplets by modifying the collision efficiency of aerosol particles and droplets. Early laboratory studies found that raindrops (of around 0.5~mm diameter) are about a factor 100 more efficient at collecting aerosol when they are charged rather than neutral \cite{barlow83}. If only one particle is charged, then the collision efficiency is always enhanced since image charges are attractive. In the case of both particles having the same charge, the effect can be either attractive or repulsive. For aerosols colliding with large cloud droplets ($>$10~$\mu$m), the effect of greater same-sign charge is to increase the collision efficiency---by up to an order of magnitude for 100$e$ charged droplets  \cite{tinsley06}. However, for aerosols or small droplets ($<$4~$\mu$m), same-sign charge leads to a strong suppression of collision efficiency---by up to several orders of magnitude for 100$e$ charged aerosols or particles.  

The proposed mechanism linking GCRs, the atmospheric electrical current and cloud particle collision efficiencies and droplet freezing rates is plausible. However, direct observations in the atmosphere of changes in cloud properties due to changes in the ionosphere-earth current have not yet been made, and are unlikely to succeed in the near future since there is only a very poor understanding at present of what controls ice formation in clouds---even in the absence of any ionisation effect. Moreover, even the \textit{sign} of a possible ionisation effect is ambiguous, since it depends on the details of the particle size and charge distributions involved.

\subsection{GCR-cloud observations}

\subsubsection{Interannual time scale}

Studies of the effects on clouds of atmospheric ions from cosmic rays extend as far back as C.T.R.\,Wilson at the beginning of the last century, whose work on simulating ion-droplet processes led to a Nobel Prize for the cloud chamber \cite{wilson12}. The cosmic ray link between clouds and solar variability was originally postulated in detail by Dickinson \cite{dickinson75} in 1975, expanding on a suggestion made by Ney \cite{ney59} in 1959. Correlations between cosmic ray changes and clouds were first reported in the 1990's by Pudovkin \cite{pudovkin97} and by Svensmark \cite{svensmark97}. Since these first observations there have been a large number of papers in the literature either disputing (e.g.\,\cite{kernthaler99, jorgensen00, kristjansson00, kristjansson02, sun02, damon04}) or supporting  (e.g.\,\cite{usoskin04,harrison06,vieira06}) the presence of significant correlations between cosmic rays and clouds.

The original GCR-cloud observations were based on cloud data from a composite of several satellites, including a relatively short time series (1983--1993) from the ISCCP (International Satellite Cloud Climatology Project) \cite{rossow96}. Since then the ISCCP data have been extended over the full period from July 1983 to June 2005, and they now constitute the longest available continuous global measurement of clouds. Almost all cosmic ray-cloud time-series analyses are now based on the ISCCP data product, so they cannot be considered as independent results.  The cloud measurements are quite challenging, involving the inter-calibration of instruments on board several geostationary satellites and polar-orbiting satellites, and controlling instrumental
 drifts to better than 1\% over a period of more than 20 years. 
Although the ISCCP low-cloud data appear to follow GCR variations well until 1994 \cite{marsh00}, they show a poor correlation after that time, which is evidence against a continuing GCR-cloud effect \cite{laut03}. However, it has been suggested that there may be a long-term trend in the cloud data due to other processes, which is then modulated by GCR variations \cite{usoskin04}. After detrending the cloud data, a reasonable consistency with GCR modulation is found over the full period and, furthermore, the GCR-cloud correlation suggests an increased amplitude at higher latitudes, consistent with what may be expected from GCRs \cite{usoskin04}. Questions have also been raised about apparent discrepancies between the ISCCP cloud measurements and the independent cloud data from the  SSMI (Special Sensor Microwave Imager) instruments on board the DMSP (Defence Meteorological Satellite Program) satellites \cite{marsh03}. The discrepancies emerge around the end of 1994, which coincides with a period of several months during which no polar-orbiting satellite was available for inter-calibration of the ISCCP geostationary satellites.

Assuming for the moment that the low cloud data do indeed show a solar imprint, then the  modulation measured for low clouds is about 1\% absolute (3\% relative), which represents around 0.5~\wpm\ variation in net radiative forcing at Earth's surface over the solar cycle---about a factor 2.5 larger than the variation of solar irradiance. Furthermore, the two effects are in phase, namely increased sunspot activity corresponds to increased irradiance, decreased GCR intensity and decreased low cloud amount. Since low clouds have a net cooling effect, this implies increased radiative forcing and a warmer climate. During the first half of the twentieth century, the cosmic ray intensity fell by about one unit of solar cycle modulation  (Fig.\,\ref{fig_be10_solarb}), so a putative GCR-cloud effect could have contributed a long-term radiative forcing during this period of around 0.5~\wpm\ , which is substantially larger than the estimated forcing due to solar irradiance changes (Fig.\,\ref{fig_solar_irradiance}). This underscores the importance of establishing whether or not cosmic rays do indeed affect clouds.

If there is such an effect, it is likely to be restricted to particular geographical locations and atmospheric conditions, such as concentrations of certain trace gases, ambient aerosol concentrations, temperature (altitude), etc. Consequently, it is more appropriate to study GCR-cloud correlations according to geographical region and altitude, rather than globally. Several regional studies have been made, e.g.\,\cite{marsh00,marsh03,palle04,voiculescu06}, and the effects of volcanoes and the El Nin\~{o} Southern Oscillation (ENSO) also accounted for \cite{marsh03,voiculescu07}. These studies find significant correlations between clouds and GCRs in certain climate-defining regions, and find different correlations depending on whether the clouds are low, medium or high. One study \cite{voiculescu06} evaluates separately the correlations of low clouds with cosmic rays and with UV intensity in order to try to discriminate between the two forcing mechanisms. However, the conclusions of all these studies are subject to the uncertainties raised above concerning the presence or absence of a trend in the cloud data, and the open question about possible calibration uncertainties. This is especially of concern when trying to distinguish between the various forcing mechanisms \cite{voiculescu06},  since all solar indices---irradiance, UV and GCRs---have quite similar variations over the solar cycle.

Overall, therefore, the present satellite observations, while suggestive, are insufficient either to establish or to rule out an effect of cosmic rays on clouds, and a further extension of the ISCCP data is required to clarify the situation.  The observations do, however, provide sufficient evidence to suggest that a deeper investigation of the effect of cosmic rays on clouds offers the real possibility of fundamental new knowledge on the solar-climate problem. 

\subsubsection{Daily time scale}

In addition to the quasi-11-year solar cycle modulation, GCRs also display sporadic short-term reductions known as Forbush decreases.  During a Forbush decrease the GCR intensity falls typically by around 5\% in the first day and then returns to normal levels over the next few ($\sim$5) days. Larger Forbush decreases may occur, with as much as 20\% reduction of the GCR intensity measured in ground-level neutron monitors. They are caused by magnetohydrodynamic disturbances in the solar wind, following large solar coronal mass ejections. If the ejection is directed towards Earth, the Forbush decrease of GCR intensity is generally more pronounced and frequently accompanied by a burst of solar protons, with characteristic energies up to around 300~MeV, which may (over-)compensate the ionisation reduction in the stratosphere and upper troposphere. Any atmospheric responses during Forbush and solar proton events can be unambiguously associated with changes of ionising particle radiation, without the usual ambiguity with solar UV changes over the solar cycle. This, together with their short duration, make Forbush events an important laboratory for testing possible cosmic ray-climate forcing.

From a superposed epoch analysis of Forbush events over the period 1986--1994, a small (1.4\% absolute) but significant reduction in global cloud cover (recorded in ISCCP data) was found during Forbush decreases that are unaccompanied by solar proton events \cite{todd01}. Significant cloud anomalies were found to be concentrated at high latitudes and most pronounced (up to 30\% reduction) in the uppermost clouds over Antarctica (at 10-180 mbar, which corresponds to polar stratospheric clouds). In contrast, Forbush decreases associated with solar proton events showed no significant cloud anomalies. All of these observations are consistent with a GCR-cloud effect in the region of the atmosphere that experiences the greatest change of ionisation during a Forbush decrease (i.e.\,high latitude and high altitude). A recent update has extended this analysis over the period 1983--2000 \cite{todd04}. Cloud anomalies are found with similar space and time structure as previously, although of smaller magnitude. Pronounced reductions of polar stratospheric clouds over Antarctica in association with Forbush decreases are also found in the updated analysis, restricted to the local winter months when polar stratospheric clouds are most prevalent. However it is pointed out that these results should be treated with caution since the absence of daylight over Antarctica during the austral winter precludes multi-spectral analysis in the ISCCP data, and so the cloud-type identification relies on infra red measurements alone. Furthermore, the Forbush events are associated with statistically significant positive temperature anomalies over Antarctica, by up to 12\degc\ at the surface and extending throughout the troposphere. This latter observation appears to be inconsistent with a simultaneous reduction of cloud cover, since Antarctic clouds should have a net warming effect due to longwave absorption and re-radiation.

A wide range of meteorological responses to short-term changes in the global electrical circuit, induced by energetic solar events or crossings of the heliospheric current sheet, have also been reported in the literature \cite{tinsley04}. They include changes in atmospheric temperature, pressure or dynamics, such as an influence of solar proton events on cyclone development in the North Atlantic \cite{veretenenko04}. The meteorological responses are generally small in magnitude but of high statistical significance. They correspond to changes in $J_z$ of around 5\% at low latitudes and up to about 25\% at high latitudes. Among the most climatically-significant observations is the dramatic deviation of wind directions reported at Antarctic stations located in the Western Pacific and near to the south pole (Vostok) in response to disturbances caused by crossings of the solar wind magnetic sector boundaries \cite{troshichev05}. The standard wind pattern in the Antarctic is cold katabatic winds that descend over the region due to negative air buoyancy and drain radially outwards along the slope of the Antarctic ice sheet. Rapid warming events have been observed of up to 20\degc\ in a few hours, which are found to be associated with a large increase of the southward (negative) component of the interplanetary magnetic field (IMF) carried by the solar wind \cite{troshichev05}. This polarity couples the IMF with the geomagnetic field and increases the ionisation and $J_z$ of the polar atmosphere. From the observed pattern of temperature changes with respect to altitude, the meteorological changes are attributed to a cloud layer appearing at 5--10 km altitude under the influence of the increased ionisation. Associated with these changes of the IMF are anomalous winds, both in terms of their direction and high speed. These seem to be associated with ENSO activity, occurring around 1--2 months in advance of ENSO, although the present statistical significance is limited. This association of solar magnetic activity with ENSO has been proposed previously, based on analysis of the Sun's motion around the centre-of-mass of the solar system \cite{landscheidt00}. 
  
Associations have also been reported between rainfall and cosmic rays. A study of the influence of Forbush decreases on precipitation in Brazil and in the former Soviet Union found a decrease of rainfall of $(17 \pm 3)$\% on the day of the Forbush decrease \cite{stozhkov01}.  On the other hand, for ground-level solar proton events (solar cosmic rays) an increase in rainfall of $(13 \pm 5)$\% was found over the same period. These observations are consistent with a separate study of the relationship of GCR flux with global precipitation over the period 1979--1999 \cite{kniveton01}. Both the precipitation and precipitation efficiency (ratio of precipitation to total columnar precipitable water) showed a positive correlation with solar-cycle GCR variations by 7--9\% over the latitudinal band 45--90$^\circ$S. Alternative explanations for the rainfall variations, such as ENSO or changes in tropospheric aerosols, showed poorer statistical relationships.

\subsection{GCR-cloud-climate mechanisms}
\label{sec_gcr_cloud_climate}

\subsubsection{Importance of aerosols and clouds}

Clouds account for a global average 28 \wpm\ net cooling \cite{hartmann93}, so changes of their coverage or reflectivity by only a few per cent can have a significant effect on the climate. This figure represents the difference between the cooling effect due to reflection of the incoming shortwave radiation, and the absorption of longwave terrestrial radiation, which warms the atmosphere and surface. Low-altitude layered clouds covering large areas of the tropics and subtropics make the greatest contribution to cooling Earth since they efficiently reflect intense incoming solar radiation, while at the same time are prodigious emitters of longwave radiation. These are the marine stratus and stratocumulus clouds---the ``climate refrigerators of the tropics'' \cite{bretherton04}.

Water vapour and clouds have an important influence on the redistribution of energy. Evaporation from the surface carries large amounts of heat away from the surface, and  condensation of the rising water vapour releases this heat in the troposphere, supplying energy to atmospheric circulation. 
Recent observations suggest that precipitation and total atmospheric water have increased at about the same rate of 7\% per K of surface warming over the past two decades, consistent with the Clausius-Clapeyron equation \cite{wentz07}. This implies that current climate models---which predict that global precipitation will increase at only 1--3\% per K \cite{wentz07}---are substantially under-estimating the transfer of latent heat from the surface to the atmosphere. Latent heat accounts for a radiative transfer of about 80~\wpm\ from Earth's surface, so a 5\% discrepancy per K implies about 2~\wpm\ under-estimate of latent heat transfer for the current warming. 

Ice formation is especially important to the energy released by clouds. Air in warm clouds is typically only a few tenths of a per cent super-saturated with respect to liquid water. However, this represents a large supersaturation with respect to ice (several tens of per cent) and, in consequence, any ice particles that are present will grow rapidly at the expense of the water droplets, and induce precipitation by sedimentation.

A strong connection between aerosol loading and cloud properties has been demonstrated both for shallow \cite{kaufman05} and deep convective \cite{koren05} clouds, using data from the MODerate resolution Imaging Spectroradiometer satellite, MODIS \cite{solomonson89}. Increased aerosol loading is associated with increased cloud fraction, decreased cloud droplet radius, increased cloud reflectivity, increased cloud water content, and increased cloud-top altitude.  The highest clouds show a constant or decreasing ice cloud optical depth with increased aerosol loading, suggesting stronger convection and the formation of higher towers with more extensive (thinner) ice anvils. These correlations occur repeatedly, and independently of the type of cloud or type of aerosol, from fine (pollution, smoke) to coarse (sea salt, terrigenous dust), although with different sensitivities. 

%---- Begin Fig. (includegraphics)  ----
\begin{figure}[htbp]
  \begin{center}
      \makebox{\includegraphics[width=125mm]{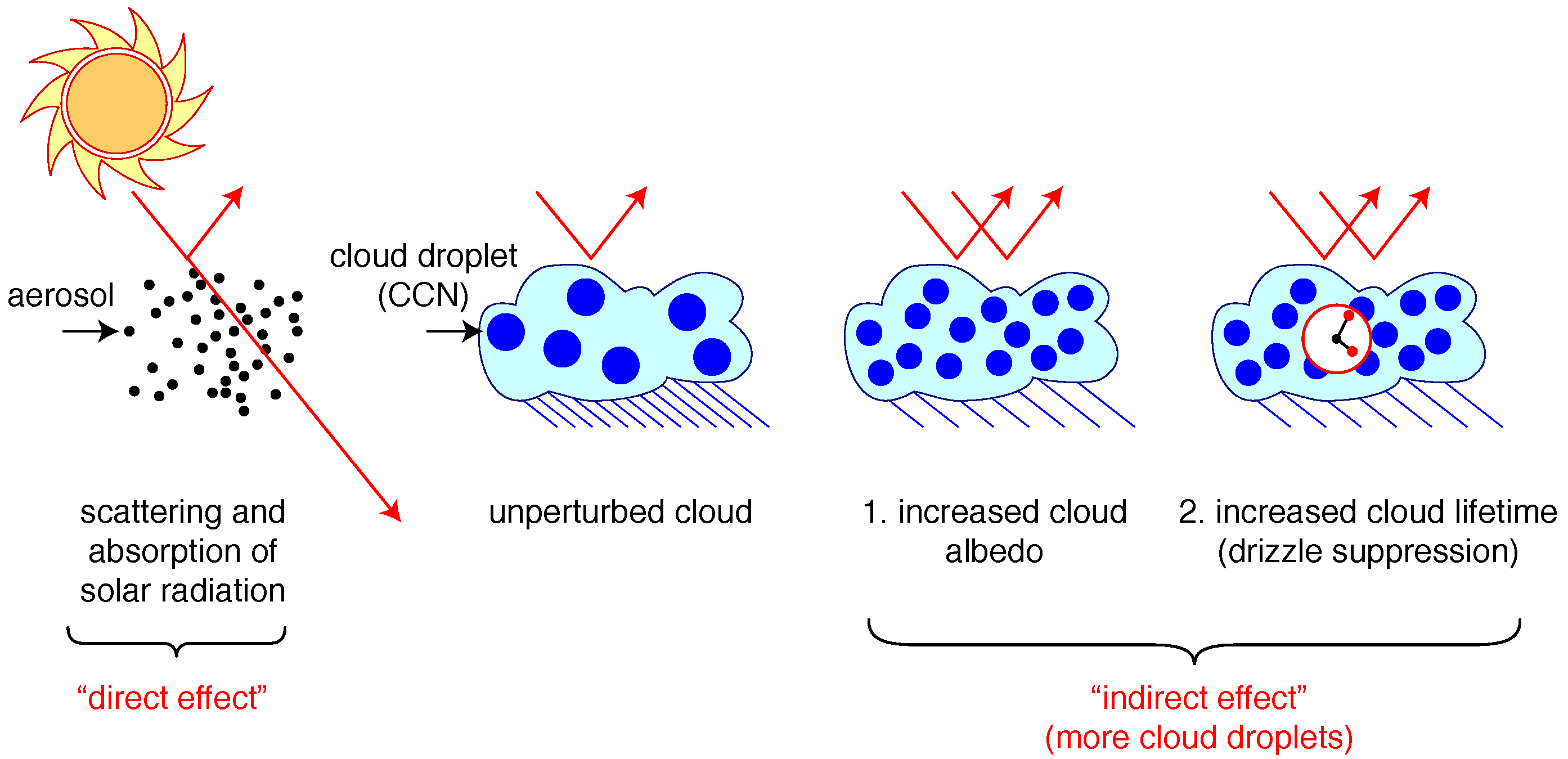}}
%      For a frame: 'makebox' => 'framebox[textwidth]'
%      For a bitmap file: '.png' => '.jpg'
  \end{center}
%  \vspace{-5mm}
  \caption{Radiative forcing due to aerosols, showing the direct effects and indirect effects (via clouds).}  
  \label{fig_aerosol_radiative_forcing} 
  \end{figure}
%---- End Fig. (includegraphics)  ----

These observations confirm that increased aerosol loading leads to increased CCN concentrations and, in turn, to clouds with a larger number of droplets, with a smaller mean size. This has important consequences on the ``indirect effects'' of aerosols on climate through their influence on radiative properties (Fig.\,\ref{fig_aerosol_radiative_forcing}). The first indirect effect is to increase the droplet surface area, making the cloud more reflective (known as the Twomey effect \cite{twomey77}). The second indirect effect is the inhibition of precipitation, which increases the thickness of the cloud and its lifetime, and hence the cloud amount.  For shallow clouds over the tropical and subtropical Atlantic Ocean, changes of the top-of-the-atmosphere radiative forcing due to increased aerosol loading are found to be largely due to the increased cloud amount (80\%), with smaller contributions from the Twomey effect (10\%) and increased liquid water path (10\%)  \cite{kaufman05}.

\subsubsection{Marine stratocumulus}

The geographical regions that are most likely to be climatically sensitive to ion induced nucleation from GCRs are those where the existing CCN concentration is very low. For such regions, small variations in aerosol concentrations can induce large changes of cloud cover. Promising candidates are the vast regions of the eastern tropical and subtropical Pacific Oceans covered by marine stratocumulus clouds \cite{rosenfeld06}.

It is common for low clouds within the narrow marine boundary layer to self-organise into Rayleigh-B\'{e}nard convective cells, limited on their upper edge by the temperature inversion. In the eastern subtropical oceans, more than 60\% of all cloud scenes viewed in satellite images are classified as so-called mesoscale cellular convection \cite{wood06}. (Ironically, Rayleigh-B\'{e}nard convection is also responsible for the plasma granulation cells covering the solar photosphere.) The Rayleigh-B\'{e}nard cells are probably initiated by radiative cooling of the top of the cloud, together with entrainment of unsaturated air parcels from the free troposphere, causing a convective downdraft by evaporative cooling \cite{randall80}. Drizzle from the convective cells continually removes aerosols and may result in open cells in which only narrow clouds exist in the regions of convective updraft at the edges.  The convective cells are only stable in narrow boundary layer clouds of a few 100 m thickness since the Rayleigh number that characterises the cells' stability varies as the fourth power of the cloud thickness; for large values of this number there is turbulent mixing between the upward- and downward-going air parcels.

A curious feature of regions where marine stratocumulus predominate is the appearance of large pockets of almost cloud-free air (open cells) embedded in regions of otherwise homogeneous cloud field (closed cells) \cite{stevens05}. These open pockets are remarkably stable, lasting of order tens of hours. It suggests that the cells have two relatively stable states: 1) closed cells in which vigorous Rayleigh-B\'{e}nard convection is maintained by entrainment from the free troposphere, drawing in fresh CCN, and 2) open cells of weaker Rayleigh-B\'{e}nard convection that prevents replenishment of CCN from the free troposphere and is therefore depleted of CCN. Open cells correspond to extreme CCN depletion; small increases in the CCN population can close these cells and induce a very large change in surface radiative energy balance for the region concerned (increase of cloud coverage from about 50\% to 100\%). This picture is supported by observations of extremely light rainfall in the closed cell regions ($<$1\% instantaneous frequency, and mostly evaporating before reaching the surface) and, on the other hand, heavy surface precipitation of about 10 mm/day from the walls of the open cells  \cite{stevens05}. The latter is highly significant since it is more than twice the mean surface evaporation rate and, without replenishment, it would deplete the cloud of all its liquid water (and CCN) in tens of minutes. The existence of extremely low CCN concentrations in the open cell regions is also supported by the presence of pronounced ship tracks due to the injection of aerosols. 

%---- Begin Fig. (includegraphics)  ----
\begin{figure}[htbp]
  \begin{center}
      \makebox{\includegraphics[width=115mm]{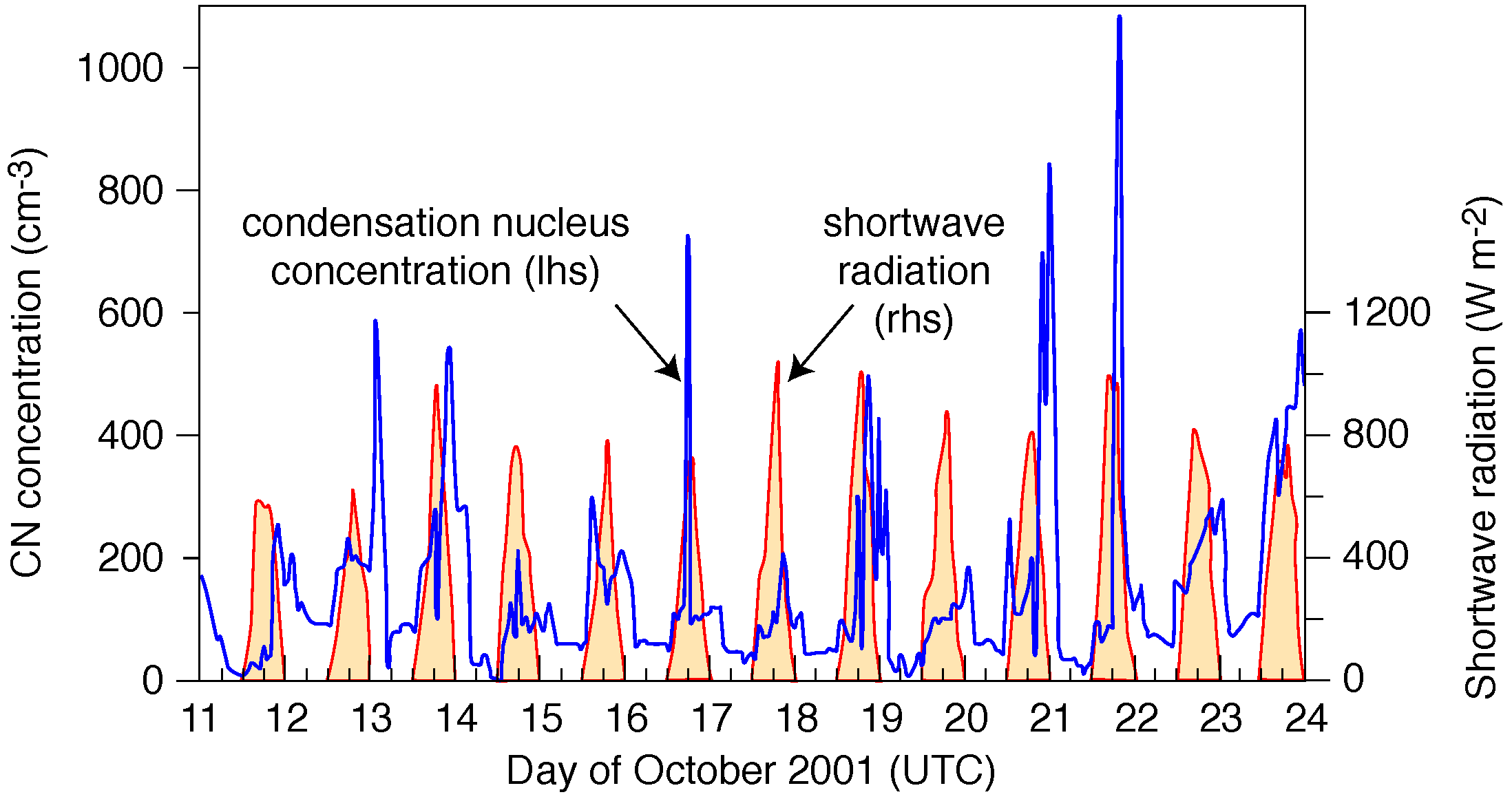}}
%      For a frame: 'makebox' => 'framebox[textwidth]'
%      For a bitmap file: '.png' => '.jpg'
  \end{center}
  \caption{Condensation nucleus (CN) concentration (10~nm diameter threshold) and shortwave radiation measured at sea surface level during the EPIC 2001 Stratocumulus Study in the south-east Pacific Ocean \cite{bretherton04}.}  
  \label{fig_aerosol_bursts_epic} 
  \end{figure}
%---- End Fig. (includegraphics)  ----

During the EPIC 2001 Stratocumulus Study in the south-east Pacific Ocean, measurements were made of the condensation nuclei (CN) concentration (10~nm diameter threshold) and shortwave radiation (SWR) at sea surface level (Fig.\,\ref{fig_aerosol_bursts_epic}) \cite{bretherton04}. A strong diurnal cycle was observed of fresh CN, driven by photochemical processes, on a background of low night-time CN concentrations that correlate well with cloud droplet concentrations. Cloud droplet concentrations of 100 cm$^{-3}$ or less in boundary layer clouds produce substantial drizzle and CCN sedimentation \cite{wood05}. Although unexplained at the time \cite{bretherton04}, these daytime aerosol bursts are consistent with the nucleation and growth of  \htwosofour-\htwoo\ aerosols. The \htwosofour\ concentrations are too low during the nighttime for nucleation to occur; however, during the day, photochemical production of OH radicals allows for the oxidation of di-methyl sulphide (DMS) to \sotwo, and then to \htwosofour. Nevertheless, sufficient concentration of \htwosofour\ for nucleation and growth to occur is frequently only reached by the latter part of the day (Fig.\,\ref{fig_aerosol_bursts_epic}). These are precisely the conditions under which ion induced nucleation could play a major role in determining the production rate of fresh aerosols and, by implication, the concentration of CCN and therefore cloud amount.

There is support for this picture from an earlier measurement of particle nucleation in the tropical boundary layer in which all of the key species, including OH, \htwosofour, \sotwo\ and DMS, were measured \cite{clarke98}. The observations unambiguously link aerosol nucleation and growth in the eastern tropical Pacific boundary layer to the marine sulphur cycle. Moreover, the measurements demonstrated that nucleation of \htwosofour\ particles is taking place at concentrations nearly an order of magnitude below those required by classical nucleation theory, which does not consider charge. On the other hand, the data were found to be well explained by a model of ion-induced nucleation \cite{yu01}.

Further support for a possible GCR-cloud effect comes from satellite cloud data, which show a strong solar cycle modulation of the radiative fluxes from low clouds in the southeastern Pacific Ocean \cite{vieira06}. This is attributed to the influence of increased GCR flux from the Southern Hemisphere Magnetic Anomaly (SHMA) \cite{vieira06}---a large region of relatively low total geomagnetic field centred on South America, around 30$^\circ$S. However, it is the horizontal field that is responsible for GCR shielding, and this component of the SHMA is centred much further to the east, near the tip of South Africa. On the other hand, the region of the southeastern Pacific Ocean dominated by mesoscale cellular convection \cite{wood06} coincides closely with the  observed region of strong solar cycle modulation. An alternative and perhaps more plausible explanation for the enhanced solar modulation is therefore that the region is especially climatically sensitive  to the effect of GCRs on new particle production---which may influence the ratio of closed vs.\,open mesoscale convective cells. In summary, there is quite suggestive evidence that marine stratocumulus clouds in the tropics and sub-tropics are good candidates for further study of a possible cosmic ray influence.

\subsubsection{Lightning and climate}
\label{sec_lightning_climate}

\begin{figure}[htbp]
  \begin{center}
      \makebox{\includegraphics[width=160mm]{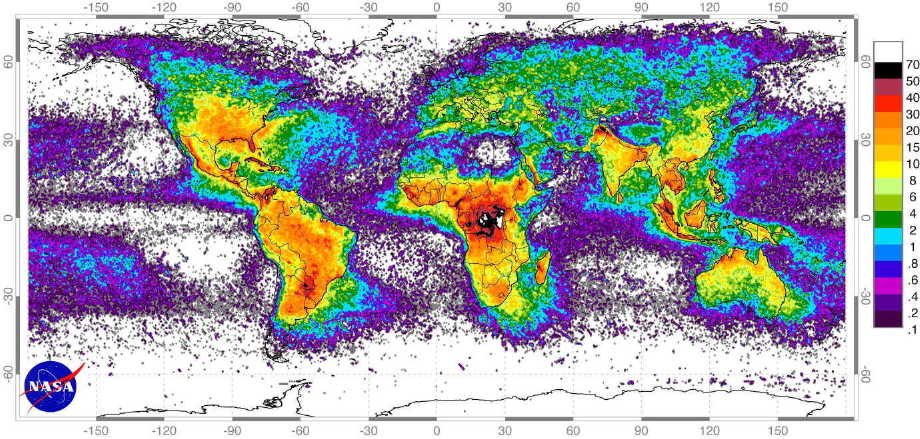}}
%      For a frame: 'makebox' => 'framebox[textwidth]'
%      For a bitmap file: '.png' => '.jpg'
  \end{center}
  \vspace{-5mm}
  \caption{Global distribution of lightning flash rate (km$^{-2}$y$^{-1}$), April 1995--February 2003, from the combined observations of the NASA Optical Transient Detector (OTD) and Lightning Image Sensor (LIS) satellite instruments. The pixel size is 0.5$^\circ \times $0.5$^\circ$.}  
  \label{fig_lightning_map} 
  \end{figure}

A great deal has been learnt in recent years about global lightning activity and the global electrical circuit from satellite observations and measurements of natural (Schumann) resonances of the Earth-ionosphere cavity \cite{williams05}. Lightning activity is an order of magnitude larger over land than over oceans, and it is concentrated in the tropics, following the Sun's zenith with the seasons (Fig.\,\ref{fig_lightning_map}). Charge separation occurs in the mixed-phase region of cumulonimbus clouds by collisions between large graupel particles and ice crystals (or small graupel particles or snow). The collisions transfer negative charge to the larger particles, which then separate gravitationally to produce intense large-scale electric dipoles. These may discharge to ground, carrying negative charge downwards and positive charge up to the ionosphere (Fig.\,\ref{fig_global_electric_circuit}).

%---- Begin Fig. (includegraphics)  ----
%---- End Fig. (includegraphics)  ----

The process of charge separation is very sensitive to cumulonimbus updraft speed. This is thought to account for the large contrast in lightning activity over land compared with the ocean. The land is more readily heated by solar radiation than the ocean, which is mobile and has a higher heat capacity. Another factor that produces more vigorous updraft speeds over land is the higher base height \cite{williams05}. Large lightning variations are observed of around 50--100\% for small temperature variations of only 1\degc. These have been observed on time scales from diurnal to that of the El-Ni\~{n}o-Southern Oscillation, ENSO (3--5 yr). Whether the strong link between lightning and global temperature persists on longer time scales is an open question since the long-term convective re-adjustment of the atmosphere is not well understood.

Is there any link between lightning variability and GCR-cloud-climate effects? The answer is yes, perhaps, in two\footnote{We do not include here the contribution of cosmic ray showers as the likely trigger for lightning, by injecting MeV-energy electrons into electrified clouds, which can then avalanche and lead to runaway breakdown.} quite distinct areas: aerosol production and Milankovitch orbital cycles.
In the case of aerosol production, it has been proposed that the large contrast in land-ocean lightning activity could be due to the increased aerosol concentrations over land \cite{williams02a}. This has the effect of producing smaller droplets and suppressing coalescence and precipitation in the warm cloud base. In this way, a larger amount of water reaches the mixed-phase region, allowing for more vigorous and abundant charge separation. Indeed, the global distribution of lightning activity (Fig.\,\ref{fig_lightning_map}) shows considerable similarities with that of the aerosol optical depth measured by satellites. However, studies of lightning activity over the Amazon \cite{williams02a} and over islands of sizes spanning ocean-like to continental-like rates \cite{williams02b,williams04} do not support the aerosol hypothesis---although they are unable to rule it out.

In the second case, it is quite plausible that lightning activity may be modulated by the Milankovitch orbital insolation cycles. This would imprint the orbital modulation frequencies on the  global electrical current and, in turn, on cloud cover, through the mechanisms described in  \S\ref{sec_global_electric_circuit}. Present low latitude insolation is about 7\% greater at perihelion (January) than aphelion (July).  Precession cycles this maximum through the seasons with a period of about 23~ky. Despite the lower insolation, lightning activity is presently 10\% larger in July than in January. This is attributed to the larger land area in the northern hemisphere compared with the southern hemisphere (Fig.\,\ref{fig_lightning_map}). In 11.5 ky, perihelion will coincide with the northern hemisphere summer and this could lead to a large increase in lightning activity, given the highly non-linear processes involved. In short, the asymmetric land masses in the northern and southern hemispheres, combined with the large land-ocean contrast of lightning activity, could provide a natural 23 ky periodicity on the ionospheric potential, $J_z$ and cloud cover. The 41~ky cycle, due to changes in obliquity of Earth's axis, would also be present in lightning activity---for the same reason of the north-south land mass asymmetry.

\subsubsection{Climate responses}
\label{sec_climate_responses}

There are several important general characteristics of a putative GCR-climate forcing. The first, which has been described in detail, is that it can act on any time scale from days to hundreds of millions of years. In particular, the GCR flux is a candidate for the presently-unknown forcing agent for centennial and millennial scale variability. Another characteristic is that a change of GCR flux acts globally and, moreover, with the same phase everywhere. Satellite data suggest that increased GCR flux is associated with increased low clouds, which globally exert a net radiative cooling effect---although a warming effect in polar regions, where clear-sky insolation is low.
A global forcing mechanism avoids the frequent puzzle of an observed synchronicity among climate transitions reconstructed at widely separated geographical locations, without any clear mechanism for their teleconnection. Orbital modulation of solar insolation also acts globally, but with an opposite phase in the two hemispheres, such that the net change of  insolation is close to zero. In contrast, a GCR-cloud forcing would cause a non-zero net insolation change, and so a relatively small GCR forcing over a long period has the potential to exert a strong climatic response. It is also important to note that, prior to industrialisation, the prevailing aerosol concentrations over most of the troposphere---including over land masses---may have been very low, and similar to present-day pristine marine environments ($\lappeq$100 CCN/cm$^3$) \cite{andreae07}. Under these conditions, cosmic rays may have had the potential to affect clouds over a larger region of the globe than today. Finally, GCR forcing of clouds would produce a rapid effect---of order one day---so all subsequent climatic responses would be observed either to be simultaneous with a GCR flux change or else to lag by their characteristic response time.

\begin{figure}[htbp]
  \begin{center}
      \makebox{\includegraphics[width=1.0\textwidth]{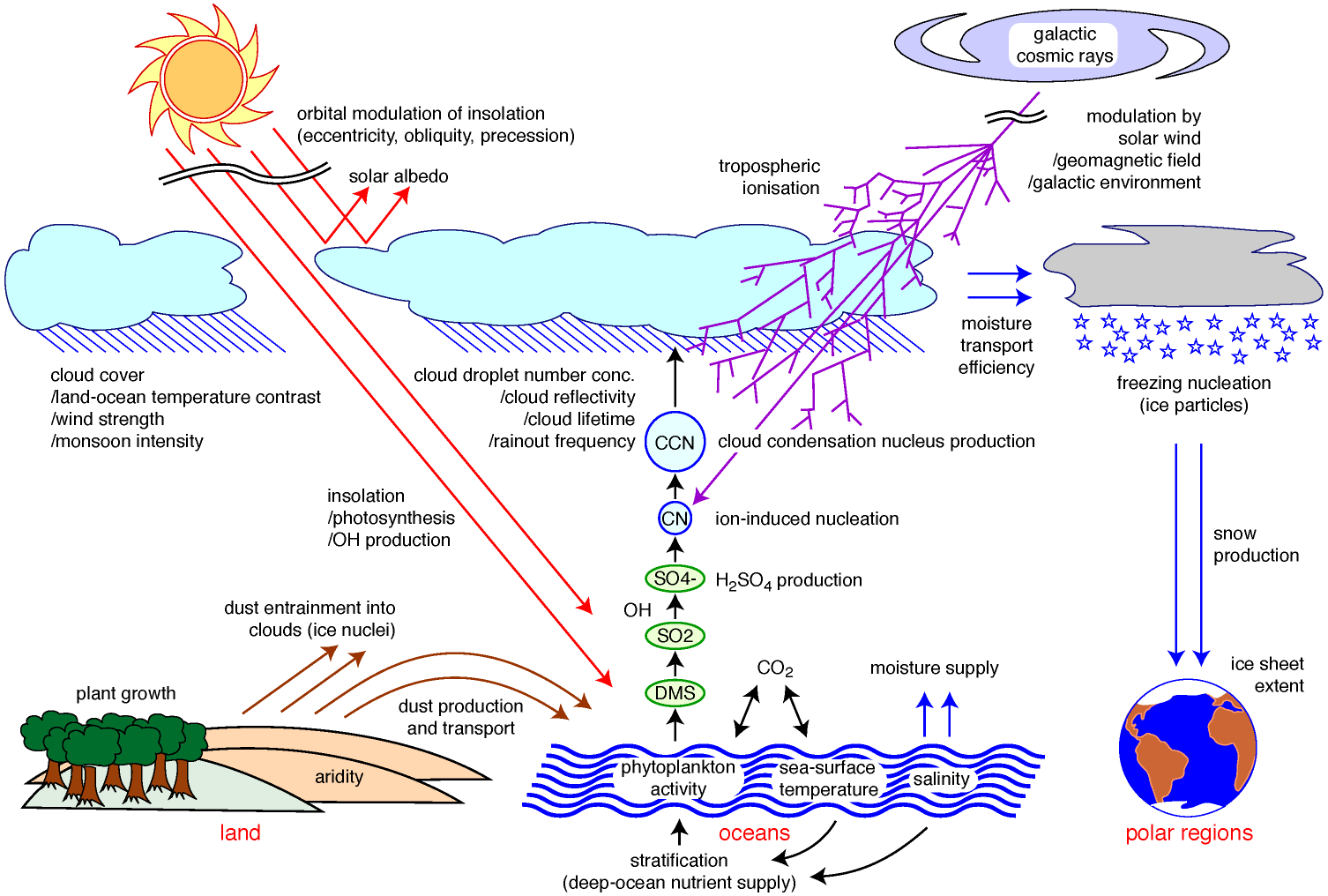}}
%      For a frame: 'makebox' => 'framebox[textwidth]'
%      For a bitmap file: '.png' => '.jpg'
  \end{center}
%  \vspace{5mm}
  \caption{Potential responses of the climate system to forcing from galactic cosmic rays, assuming they influence cloud cover and droplet number concentration. This would affect both insolation and the hydrological cycle which, in turn, may produce a wide range of climate responses, as indicated.}  
  \label{fig_gcr_climate_mechanisms} 
  \end{figure}

At present clouds are considered solely to be part of the response of the climate system to other changes, such as an increase in temperature or humidity. However, if a significant GCR-cloud interaction exists, this would introduce a new role for clouds as a primary driver of climate change. We can consider how this would influence climate (Fig.\,\ref{fig_gcr_climate_mechanisms}). 
Via some mechanism---such as ion-induced nucleation of \htwosofour\ vapour---the GCR flux is assumed to modulate the concentration of CCN and thereby cloud amount, reflectivity and precipitation. Increased GCR flux implies increased CCN concentrations, longer cloud lifetimes and higher albedo, resulting in lower insolation and more efficient long-range transport of moisture. GCR-driven variations of clouds and insolation would in turn cause feedback responses from a wide range of climatic variables which would mostly amplify the climate transition. The most important responses may include the following: 1) land and sea surface temperatures and temperature contrasts, which will affect wind and monsoon intensity, 2) moisture transport efficiency, which will affect long-wave absorption, the availability of water vapour for clouds and the ice sheet extent during the dry glacial periods, 3), aridity and vegetation, which will affect the dust supply and, in turn, phytoplankton activity, and therefore atmospheric concentrations of \cotwo.

%---- Begin Fig. (includegraphics)  ----
%---- End Fig. (includegraphics)  ----

We may speculate how the GCR flux could affect the ITCZ migration and monsoon intensity. The  monsoon system appears to be sensitive to quite small insolation changes, since monsoons around the entire tropics seem to have intensified and weakened over the last 50 years in step with the 11-year solar cycle \cite{vanloon04}. The sign of the effect is the same as that observed in the palaeoclimatic reconstructions, namely a decreased  GCR intensity is associated with a strengthening of the monsoon (increased rainfall). Anomalies in the seasonal migration of the ITCZ can occur due to shifts in the north-south gradient of ocean surface temperatures near the equator \cite{chiang04}. Following the earlier discussion, the mesoscale convective cells in subtropical stratocumulus clouds may be especially sensitive to ion induced nucleation, and so any north-south asymmetry of these clouds could produce a corresponding ocean surface temperature gradient that depends on the GCR flux.  The ocean temperature gradient can, in turn, drive a low-level atmospheric circulation that may strengthen the trade winds in, for example, the south and weaken them in the north. The result is to impede the southward migration of the ITCZ, increasing northern tropical rainfall and reducing southern tropical rainfall. This condition would  persist for the duration of the forcing anomaly which, for GCRs, may last several centuries or more. 

Among the consequences of GCR-induced effects on the hydrological cycle and wind strength, potential changes in dust production and transport may be especially important for glacial-interglacial climate differences. The GISP2 Greenland ice core has provided evidence of solar/GCR forcing of the dust flux during the last glacial period, to at least 100 ky ago: measurements of the dust concentration in six core sections, each representing 1--2 ky, have revealed distinct 11 y and 22 y modulations, characteristic of the solar cycle \cite{ram97}. Increased GCR flux would increase CCN concentrations and cloud lifetimes, producing decreased rainout frequency and therefore increased aridity and dust production and, in addition, an increased efficiency for transporting water evaporated from the oceans, deep into continental regions.  Since terrigenous dust is the major source of iron for the sea surface, its availability controls phytoplankton activity in so-called HNLC (high nutrient low chlorophyll) polar ocean regions. Therefore dust-supply strongly influences DMS (di-methyl sulphide) production by phytoplankton; up to eight-fold increases in DMS concentrations have been seen in iron-addition experiments \cite{turner04}. Di-methyl sulphide is oxidised to \sotwo\ and then to \htwosofour\ by the hydroxyl radical, OH. The hydroxyl radical is generated by solar ultra violet radiation \cite{rohrer06}.  Higher insolation also stimulates phytoplankton activity and therefore raises DMS production. 
As a result, in regions where ion induced nucleation may be important for CCN production, increased insolation may, paradoxically, result in \textit{reduced} sea surface temperatures if the increased cloud cover overwhelms the increase of shortwave radiation. This is a possible interpretation of the interesting observation that the strong 41-ky obliquity signal in the eastern equatorial Pacific ocean surface temperatures between 1.8 and 1.2~My ago varied in the \textit{opposite} sense to local annual insolation \cite{liu04}.

\section{CLOUD FACILITY AT CERN}
\label{sec_cloud_facility}

\subsection{Overview}
\label{sec_overview}

The key to further progress on the cosmic ray-cloud-climate question is to understand the nature of the physical mechanism. This requires a thorough experimental study of the fundamental microphysical interactions between cosmic rays and clouds.  Given the many sources of variability in the atmosphere---and the lack of any control of the GCR flux---demonstrating and explaining overall cause and effect beginning with changes in ionisation rate and ending with observations of perturbed clouds will be quite challenging. For this reason, an experimental facility known as CLOUD (Cosmics Leaving OUtdoor Droplets) is being set up at CERN to investigate GCR-cloud microphysics under controlled conditions in the laboratory \cite{cloud_proposal,kirkby02}.

\begin{figure}[tbp]
  \begin{center}
      \makebox{\includegraphics[width=140mm]{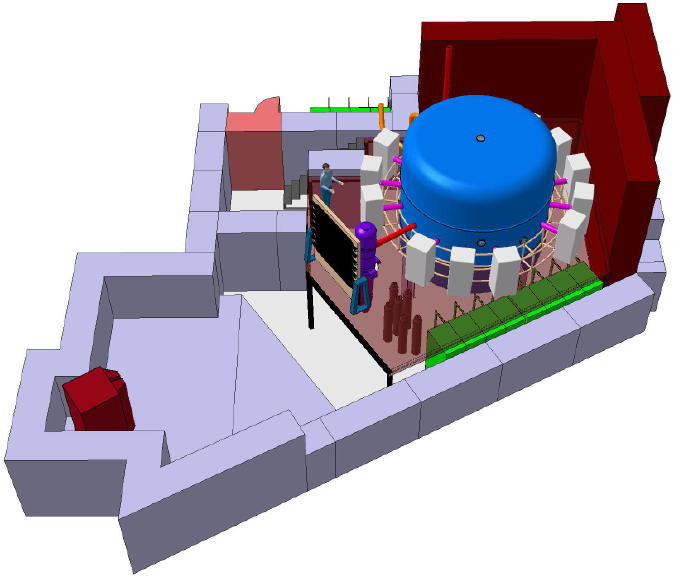}}
%      For a frame: 'makebox' => 'framebox[textwidth]'
%      For a bitmap file: '.png' => '.jpg'
  \end{center}
  \vspace{-5mm}
  \caption{Perspective view of the CLOUD facility at CERN, showing the 4 m aerosol chamber and 0.5m cloud chamber. The aerosol chamber is surrounded by instruments that sample and analyse its contents. The particle beam emerges from the dipole magnet at the lower left of the drawing and then traverses a beam counter array and the chambers. The entire experimental area is surrounded by concrete and steel radiation shielding blocks.}  
  \label{fig_cloud_4m_3d}  
  \end{figure}

The experiment involves a 4~m diameter aerosol chamber and a 0.5~m cloud chamber (Fig.\,\ref{fig_cloud_4m_3d}) which are exposed to a CERN particle beam, providing an adjustable source of ``cosmic rays'' that closely simulates GCRs at any altitude or latitude. The chambers are filled with ultra-pure air, water vapour and selected trace gases and aerosols, and can be operated at any temperature or pressure found in the atmosphere. In view of the highly non-linear nature of the processes under study, all experimental parameters (cosmic ray intensity, trace gas concentrations, temperature, pressure, etc.) will duplicate atmospheric conditions. Since the trace gas concentrations may be as low as parts-per-trillion, extremely careful attention is required for the preparation of the inner surfaces of the chambers and for the design of the gas supply system. Each chamber is equipped with UV illumination (for photolytic reactions) and an electric field cage to control the drift of small ions and charged aerosols.  During beam exposure, the contents of the chambers are continuously analysed by a wide range of sensitive instruments connected to the chambers via sampling probes, optical ports and electrical feed-throughs.

The CLOUD facility is presently in its design and prototyping phase. A prototype 3~m aerosol chamber---limited to operation at room temperature and 1 atmosphere pressure---will start operation at the beginning of 2009. The final 4~m aerosol chamber, operating at any temperature or pressure, is expected to start in 2011.

%---- Begin Fig. (includegraphics)  ----
%---- End Fig. (includegraphics)  ----

\subsection{Experimental goals}
\label{sec_experimental_goals}

The essential experimental goal of CLOUD is to carry out definitive laboratory measurements of the fundamental physical and chemical processes through which cosmic rays may affect the number of cloud droplets or ice particles. Where effects are found in the experimental data, they will first be evaluated with cloud models and, if climatically significant, parametrised for inclusion in General Circulation Models (GCMs). The CLOUD facility is designed to have sufficient flexibility to make a broad experimental survey of this poorly-known field. A series of experiments will be carried out over a period of several years to investigate the influence of cosmic rays on a) aerosol nucleation and growth (\S\ref{sec_iin}), and b) cloud microphysics such as ice particle formation and collision efficiencies between aerosol particles and cloud droplets (\S\ref{sec_global_electric_circuit}). The main experiments are outlined below.

\subsubsection{Aerosol nucleation and growth experiments}
\label{sec_aerosol_experiments}

\paragraph{Ion-induced nucleation:}

The effect of ionising particle radiation on the formation rate of ultrafine condensation nuclei (few-nm size range) from trace precursor vapours will be measured. The trace gases include, in particular, H$_2$SO$_4$, HNO$_3$, NH$_3$ and volatile organic compounds (VOCs), all in the presence of H$_2$O vapour.   The basic parameter to be measured is the nucleation rate, $J$ (\pccps), as a function of the primary experimental variables: trace vapour types and concentrations, temperature, background aerosol concentration, ion-pair production rate and aerosol charge sign.
Measurements will also be taken with electric fields of up to 20~kV/m in the chamber. This sweeps all small ions out of the chamber in about one second, effectively ``switching off'' all ionisation, including that due to natural GCRs. A typical run under fixed conditions will last several hours, after which the contents of the chamber will be replaced and the run conditions---such as particle beam intensity---adjusted in preparation for a new run.

\paragraph{Growth of CN into CCN:}

The effect of cosmic rays on the growth of CN  into CCN (i.e. from $\sim$5\,nm diameter to $\sim$100\,nm) will be measured in the presence of condensable vapours that are  known to be important for aerosol growth in the atmosphere, in particular VOCs. The main parameter to be measured at the end of each run is the CCN concentration as a function of water vapour saturation. Prior to operation of the 0.5~m cloud chamber, the CCN measurements will be made with a small sampling expansion chamber. The fraction of the CN that constitute CCN depends not only on their size but also on their chemical composition and other properties.  The evolution of the CN size spectra during beam exposure will also be measured. As before, a series of runs will be performed to take measurements as a function of the primary experimental variables: trace vapour types and concentrations, temperature, background aerosol concentration and ion-pair production rate. A typical single run may last up to about 1 day, reflecting the relatively slow growth times involved. This is the primary reason for the large size of the aerosol chamber, both to minimise the wall losses of trace vapours and aerosols, and to provide a large parent volume for the external sampling analysers.

\paragraph{Activation of CCN into cloud droplets:}

The goal of these experiments is to see how the presence of electrical charge affects the critical supersaturations required to activate CCN, and also how it affects the number of cloud droplets that appear. Well-defined aerosol sizes will be produced with standard aerosol generation techniques. The simplest systems will involve typical hygroscopic aerosols found in the atmosphere, e.g. NaCl, (NH$_4$)$_2$SO$_4$ and H$_2$SO$_4$, and also partially soluble aerosols such as CaSO$_4$. Other measurements will be made on hydrophobic carbonaceous particles, with and without the presence of organic surfactants.

\subsubsection{Cloud microphysics experiments} 
\label{sec_cloud_microphysics_experiments}

\paragraph{Ice particle formation:}

These experiments will study the effect of highly-charged aerosols $(\!\gappeq$100~$e$) on the formation of ice particles in clouds. A well-defined polydisperse distribution of CCN will first be introduced into the chamber, together with a small population  (10$^{-4}$--10$^{-6}$ relative to the CCN) of efficient ice nuclei such as AgI or ice-nucleating bacteria. The CCN will be exposed to a unipolar space charge in the chamber---created by the particle beam and a ``valley'' electric potential configuration for the field cage---to create a highly-charged aerosol population. A supercooled cloud at temperatures between 0\degc\ and -40\degc\ will then be created by an adiabatic pressure reduction in the chamber, followed by growth of droplets. The temperature of the droplets can be controlled by the initial temperature of the chamber and by the pressure drop. In the aerosol chamber, this will be achieved by opening the chamber to a small semi-evacuated auxiliary chamber; in the cloud chamber, by a piston expansion. After forming the cloud of supercooled droplets, the production of ice particles will be measured via both optical ports and sampling instruments. The presence of an ice crystal in a cloud of liquid droplets is readily identified by the CCD cameras since an ice crystal scatters much more light than a liquid droplet.  The ice fraction will also be measured with a backscattered polarised laser beam. The experiments will be repeated for various initial charge states of the aerosols in order to determine the influence of charge on the production of ice particles by contact nucleation.

\paragraph{Collision efficiencies of aerosols and droplets:}

These experiments will study the effect of highly-charged aerosols and cloud droplets on collision efficiencies in clouds. As before, a well-defined polydisperse distribution of CN and CCN will be introduced into the chamber, and then exposed to the particle beam and a unipolar space charge, creating a distribution of highly charged aerosols. The CCN population will then be activated into a polydisperse distribution of cloud droplets and the coalescence efficiencies of droplets with aerosols measured. The latter will involve both direct sampling measurements and also re-compression of the chamber to evaporate the droplets and re-measure the CN and CCN spectra after cloud processing.

\paragraph{Freezing mechanism of polar stratospheric clouds:}

These experiments will investigate the deposition freezing nucleation of \hnothree\ and water vapours onto ion clusters, forming nitric acid hydrates. Particles composed of such hydrates are the principal component of the  polar stratospheric clouds that initiate the destruction of ozone. The operating temperatures are typical polar stratospheric values of between 190 and 200~K. Nitric acid and water vapour will be present in the chamber at partial pressures representative of the stratosphere (10$^{-4}$ Pa for \hnothree\ vapour and $5 \cdot 10^{-2}$ Pa for \htwoo\ vapour, corresponding to about 10 ppb and 5 ppm, respectively). At these pressures and temperatures the nitric acid hydrates become supersaturated and can condense as crystals provided a suitable ice nucleus is present.  Sulphuric acid vapour will be included in the air mixture to represent the species most likely to contribute to the initial ion cluster formation.

\section{CONCLUSIONS}
\label{sec_conclusions}

Numerous palaeoclimatic observations, covering a wide range of time scales, suggest that galactic cosmic ray variability is associated with climate change. The quality and diversity of the observations make it difficult to dismiss them  merely as chance associations.  But is the GCR flux directly affecting the climate or merely acting as a proxy for variations of the solar irradiance or a spectral component such as UV? Here, there is some palaeoclimatic evidence for associations of the climate with geomagnetic and galactic modulations of the GCR flux, which, if confirmed, point to a direct GCR-climate forcing. Moreover, numerous studies have reported meteorological responses to short-term changes of cosmic rays or the global electrical current, which are unambiguously associated with ionising particle radiation.

Cosmic ray forcing of the climate could in principle operate on all time scales from days to hundreds of millions of years, reflecting the characteristic time scales for changes in the Sun's magnetic activity, Earth's magnetic field, and the galactic environment of the solar system. Moreover the climate forcing would act simultaneously, and with the same sign, across the globe. This would both allow a large climatic response from a relatively small forcing and also give rise to simultaneous regional climate responses without any clear teleconnection path.  
The most persuasive palaeoclimatic evidence for solar/GCR forcing involves sub-orbital (centennial and millennial) climate variability over the Holocene, for which there is no established forcing agent at present. Increased GCR flux appears to be associated with a cooler climate, a southerly shift of the ITCZ (Inter Tropical Convergence Zone) and a weakening of the monsoon; and decreased GCR flux is associated with a warmer climate, a northerly shift of the ITCZ and a strengthening of the monsoon (increased rainfall). The influence on the ITCZ may imply significant changes of upper tropospheric water vapour in the tropics and sub-tropics, potentially affecting both long-wave absorption and the availability of water vapour for cirrus clouds.

The most likely mechanism for a putative GCR-climate forcing is an influence of ionisation on clouds, as suggested by satellite observations and supported by theoretical and modelling studies. The satellite data suggest that decreased GCR flux is associated with decreased low altitude  clouds, which are known to exert globally a net radiative cooling effect. Studies of Forbush decreases and solar proton events further suggest that decreased GCR flux may reduce high altitude (polar stratospheric) clouds in the Antarctic.  Candidate microphysical processes include ion-induced nucleation of new aerosols from trace condensable vapours, and the formation of relatively highly charged aerosols and cloud droplets at cloud boundaries, which may enhance the formation of ice particles in clouds and affect the collision efficiencies of aerosols with cloud droplets. Although recent observations support the presence of ion-induced nucleation of new aerosols in the atmosphere, the possible contribution of such new particles to changes in the number of cloud condensation nuclei remains an open question. Furthermore, the parts of the globe and atmosphere that would be expected to be the most climatically sensitive to such processes are unknown, although they are likely to involve regions of low existing CCN concentrations.

Despite these uncertainties, the question of whether, and to what extent, the climate is influenced by solar and cosmic ray variability remains central to our understanding of the anthropogenic contribution to present climate change. Real progress on the cosmic ray-climate question will require a physical mechanism to be established, or else ruled out. With new experiments planned or underway, such as the CLOUD facility at CERN, there are good prospects that we will have some firm answers to this question within the next few years.

\paragraph{Acknowledgements:} I warmly thank my colleagues in the CLOUD collaboration for many stimulating discussions. I would also like to acknowledge Daniel Rosenfeld for suggesting the possible importance of cosmic rays for marine stratocumulus clouds. Finally I thank two anonymous referees for their helpful comments on the paper.

\end{document}